\documentclass[12pt]{article}
\usepackage{amssymb}
\usepackage{epsfig}
\usepackage{float}
\usepackage{graphicx}
\usepackage{amsmath}
\newcommand{\nua}[1]{\ensuremath{\rlap
           {\kern-2.5pt\ensuremath
           {\overset{\scriptscriptstyle(-)}{\phantom{\nu}}}}
           {\ensuremath{{\nu}_{#1}}}}}

\begin{document}

\begin{center}
{\bf Neutrino in  Standard Model and beyond}
\end{center}
\begin{center}
S. M. Bilenky
\end{center}

\begin{center}

{\em  Joint Institute for Nuclear
Research, Dubna, R-141980, Russia\\}
\end{center}
\begin{abstract}
After discovery of the Higgs boson at CERN  the Standard Model acquired a status of the theory of the elementary particles in the electroweak range (up to about 300 GeV).  What general conclusions can be inferred from the Standard Model? It looks that the Standard Model  teaches us that in the framework of such general principles as local gauge symmetry, unification of weak and electromagnetic interactions and Brout-Englert-Higgs  spontaneous breaking of the electroweak symmetry nature chooses the simplest possibilities. Two-component left-handed  massless neutrino fields play crucial role in the determination of the charged current structure of the Standard Model. The absence of the right-handed neutrino fields in the Standard Model is the simplest, most economical possibility. In such a scenario Majorana mass term is the only possibility for neutrinos to be massive and mixed. Such mass term is generated by the lepton-number violating  Weinberg effective Lagrangian. In this approach three Majorana neutrino masses are suppressed with respect to the masses of other fundamental fermions by the ratio of the electroweak scale and a scale of a lepton-number violating physics. The discovery of the neutrinoless double $\beta$-decay and absence of transitions of flavor neutrinos into sterile states would be  evidence in favor of the minimal scenario we advocate here.
\end{abstract}

\section{Introduction}
The discovery of neutrino oscillations in the atmospheric Super-Kamiokande  experiment \cite{Fukuda:1998mi} in the SNO \cite{Ahmad:2002jz} and other solar neutrino experiments \cite{Cleveland:1998nv,Altmann:2005ix,Abdurashitov:2002nt} and in the long-baseline reactor  KamLAND experiment \cite{Araki:2004mb}
is one of the most important recent discovery in the particle physics. The phenomenon of the neutrino oscillations was further investigated in the long-baseline accelerator K2K \cite{Ahn:2006zza}, MINOS \cite{Adamson:2013whj} and $T2K$ \cite{Abe:2014ugx} experiments
in the reactor experiments Daya Bay \cite{An:2013zwz}, RENO \cite{Ahn:2012nd} and Double Chooz \cite{Abe:2013sxa} and in the solar BOREXINO experiment \cite{Bellini:2014uqa}.

Neutrino oscillation results imply that flavor neutrino fields $\nu_{lL}(x)$ ($l=e,\mu,\tau$) are "mixtures" of the left-handed components of the fields of neutrinos with definite masses
\begin{equation}\label{numix}
\nu_{lL}(x)=\sum^{3}_{i=1}U_{li}\nu_{iL}(x).
\end{equation}
Here $U$ is unitary PMNS mixing matrix \cite{Pontecorvo:1957cp,Pontecorvo:1957qd,Maki:1962mu} and $\nu_{i}(x)$ is the field of neutrino (Majorana or Dirac) with mass $m_{i}$. Flavor fields  $\nu_{lL}(x)$  enter into Standard Model charged current (CC)
\begin{equation}\label{1CCur}
 \mathcal{L^{CC}_{I}}(x)=-\frac{g}{2\sqrt{2}}j_{\alpha}^{CC}(x)W^{\alpha}(x)+
\mathrm{h.c.}
 \end{equation}
and neutral current (NC) interactions
\begin{equation}\label{1NCur}
\mathcal{L^{NC}_{I}}(x)=-\frac{g}{2\cos\theta_{W}}j_{\alpha}^{NC}(x)Z^{\alpha}(x).
\end{equation}
Here
\begin{equation}\label{2CCur}
j_{\alpha}^{CC}(x)=2\, \sum_{l=e,\mu,\tau}\bar \nu_{l L}(x) \,\gamma_{\alpha}\, l_{L}(x)
\end{equation}
is the leptonic CC and
\begin{equation}\label{2NCur}
j_{\alpha}^{NC}(x)= \sum_{l=e,\mu,\tau}\bar \nu_{l L}(x) \,\gamma_{\alpha}\, \nu_{lL}(x)
\end{equation}
is the neutrino NC, $W^{\alpha}(x)$ and $Z^{\alpha}(x)$  are fields of $W^{\pm}$ and $Z^{0}$ vector bosons, $g$ is the electroweak interaction constant and $\theta_{W}$ is the weak (Weinberg) angle.

 We will consider now briefly {\em phenomenon of neutrino oscillations in vacuum} (see, for example, reviews \cite{Bilenky:1987ty,Bilenky:1998dt}). In the mixing relation (\ref{numix})
quantum fields enter. What about states of the flavor neutrinos $\nu_{e},\nu_{\mu},\nu_{\tau}$ in the case of neutrino mixing?

Flavor neutrino $\nu_{l}$  is produced in CC weak decays together with $l^{+}$  or
 produces $l^{-}$ in CC neutrino processes (for example, muon neutrino
$\nu_{\mu}$ is produced in the decay $\pi^{+}\to \mu^{+}+\nu_{\mu}$ or  produces  $\mu^{-}$ in the process $\nu_{\mu}+N \to \mu^{-}+X$, etc.).

From Heisenberg uncertainty relation  follows that in neutrino production and detection processes it is impossible to reveal small neutrino mass-squared differences. The state of the flavor neutrino  $\nu_{l}$ is {\em a coherent superposition of states of neutrinos with definite masses} (see, for example, \cite{Bilenky:2011pk})
\begin{equation}\label{mixrel1}
|\nu_{l}\rangle=\sum_{i}U^{*}_{li}~|\nu_{i}\rangle.
\end{equation}
Here $|\nu_{i}\rangle$ is the state of neutrino with mass $m_{i}$, momentum $\vec{p}$ and energy $E_{i}=\sqrt{p^{2}+m^{2}_{i}}\simeq p +\frac{m^{2}_{i}}{2E}$.

Small neutrino mass-squared differences can be revealed in neutrino experiments with large distances between a source and detector.  For the evolution of the flavor neutrino state we have
\begin{equation}\label{mixrel2}
|\nu_{l}\rangle_{t}=e^{-iH_{0}t}|\nu_{l}\rangle=
\sum_{i}|\nu_{i}\rangle e^{-iE_{i}t}~U^{*}_{li}=\sum_{l'}|\nu_{l'}\rangle(\sum_{i} U_{l'i}~e^{-iE_{i}~t}~U^{*}_{li})
\end{equation}
From (\ref{mixrel2}) for the probability of  $\nu_{l}\to \nu_{l'}$ transition we find the following expressions
\begin{equation}\label{mixrel3}
 P(\nu_{l}\to\nu_{l'})= |\delta_{l'l}+ \sum_{i\neq p} U_{l'i}~(e^{-i(E_{i}-E_{p})t}-1)~U^{*}_{li}|^{2}
\end{equation}
where $p$ is an arbitrary fixed index.

For the ultra relativistic neutrino we have $t\simeq L$, where $L$ is the distance between a neutrino source and a neutrino detector. From (\ref{mixrel3}) it follows that neutrino oscillations can be observed if
\begin{equation}\label{mixrel4}
(E_{i}-E_{p})~t\simeq \frac{\Delta m_{pi}^{2}L}{2E}\gtrsim 1,
\end{equation}
where $\Delta m_{pi}^{2}= m_{i}^{2}- m_{p}^{2}$. The inequality (\ref{mixrel4}) is the time-energy uncertainly relation applied to neutrino oscillations (see \cite{Bilenky:2011pk}).

In  more general case of the mixing of  three flavor neutrino fields and $n_{s}$ sterile neutrino fields $\nu_{sL}$ we have
\begin{equation}\label{numixas}
\nu_{\alpha L}(x)=\sum_{i=1}^{3+n_{s}}U_{\alpha i}\nu_{iL}(x), \quad \alpha =e,\mu,\tau, s_{1},...s_{n_{s}}.
\end{equation}
For $\nu_{\alpha}\to \nu_{\alpha'}$ ($\bar\nu_{\alpha}\to \bar\nu_{\alpha'}$) transition probability we find the following expression \cite{Bilenky:2012zp}
\begin{eqnarray}
P(\nua{\alpha}\to \nua{\alpha'})
=\delta_{\alpha' \alpha }
-4\sum_{i}|U_{\alpha i}|^{2}(\delta_{\alpha' \alpha } - |U_{\alpha' i}|^{2})\sin^{2}\Delta_{pi}\nonumber\\
+8~\sum_{i>k}\mathrm{Re}~U_{\alpha' i}U^{*}_{\alpha i}U^{*}_{\alpha'
k}U_{\alpha k}\cos(\Delta_{pi}-\Delta_{pk})\sin\Delta_{pi}\sin\Delta_{pk}\nonumber\\
\pm 8~\sum_{i>k}\mathrm{Im}~U_{\alpha' i}U^{*}_{\alpha i}U^{*}_{\alpha'
k}U_{\alpha k}\sin(\Delta_{pi}-\Delta_{pk})\sin\Delta_{pi}\sin\Delta_{pk}.
\label{Genexp4}
\end{eqnarray}
Here $\Delta_{pk}=\frac{\Delta m^{2}_{pk}L}{4E}$, $\Delta m^{2}_{ik}=m^{2}_{k}-m^{2}_{i}$ and $\alpha,\alpha'=e,\mu,\tau,s_{1},...s_{n_{s}}$.

Existing neutrino oscillation data are perfectly described if we assume three-neutrino mixing.
Two neutrino mass spectra are possible in this case:
\begin{enumerate}
  \item Normal Spectrum (NS)
\begin{equation}\label{NS}
 m_{1}<m_{2}<m_{3},\quad (\Delta m^{2}_{12}\equiv\Delta
m^{2}_{S})\ll (\Delta m^{2}_{23}\equiv\Delta
m^{2}_{A}).
\end{equation}
\item Inverted Spectrum (IS)
\begin{equation}\label{IS}
m_{3}<m_{1}<m_{2},\quad (\Delta m^{2}_{12}\equiv\Delta
m^{2}_{S})\ll (|\Delta m^{2}_{13}|\equiv\Delta
m^{2}_{A}).
\end{equation}
\end{enumerate}

For the normal neutrino mass spectrum from (\ref{Genexp4}) we find the following expression
\begin{eqnarray}
&&P^{NS}(\nua{l}\to \nua{l'})
=\delta_{l' l }
-4~|U_{l 1}|^{2}(\delta_{l' l} - |U_{l' 1}|^{2})\sin^{2}\Delta_{S}\nonumber\\&&-4~|U_{l 3}|^{2}(\delta_{l' l} - |U_{l' 3}|^{2})\sin^{2}\Delta_{A}
-8~\mathrm{Re}~U_{l' 3}U^{*}_{l 3}U^{*}_{l'
1}U_{l 1}\cos(\Delta_{A}+\Delta_{S})\sin\Delta_{A}\sin\Delta_{S}\nonumber\\
&&\mp 8~\mathrm{Im}~U_{l' 3}U^{*}_{l 3}U^{*}_{l'
1}U_{l 1}\sin(\Delta_{A}+\Delta_{S})\sin\Delta_{A}\sin\Delta_{S},
\label{Genexp5}
\end{eqnarray}
where solar and atmospheric neutrino mass-squared differences $\Delta m_{S}^{2}$ and  $\Delta m_{A}^{2}$ are determined by the relation (\ref{NS})  and we choose $p=2$.

In the case of the inverted mass spectrum we choose $p=1$. For the transition probability from (\ref{Genexp4}) we have
\begin{eqnarray}
&&P^{IS}(\nua{l}\to \nua{l'})
=\delta_{l' l }
-4~|U_{l 2}|^{2}(\delta_{l' l } - |U_{l' 2}|^{2})\sin^{2}\Delta_{S}\nonumber\\&&-4~|U_{l 3}|^{2}(\delta_{l' l} - |U_{l' 3}|^{2})\sin^{2}\Delta_{A}
-8~\mathrm{Re}~U_{l' 3}U^{*}_{l 3}U^{*}_{l'
2}U_{l 2}\cos(\Delta_{A}+\Delta_{S})\sin\Delta_{A}\sin\Delta_{S}\nonumber\\
&&\pm 8~\mathrm{Im}~U_{l' 3}U^{*}_{l 3}U^{*}_{l'
2}U_{l 2}\sin(\Delta_{A}+\Delta_{S})\sin\Delta_{A}\sin\Delta_{S}
\label{Genexp6}
\end{eqnarray}
where  $\Delta_{S}$ and  $\Delta_{A}$ are determined by (\ref{IS}).

If neutrinos with definite masses $\nu_{i}$ are Dirac particles the $3\times3$ PMNS  mixing matrix is characterized by three mixing angles and one $CP$ phase. In the standard parametrization it has the following form
\begin{eqnarray}
U^{D}=\left(\begin{array}{ccc}c_{13}c_{12}&c_{13}s_{12}&s_{13}e^{-i\delta}\\
-c_{23}s_{12}-s_{23}c_{12}s_{13}e^{i\delta}&
c_{23}c_{12}-s_{23}s_{12}s_{13}e^{i\delta}&c_{13}s_{23}\\
s_{23}s_{12}-c_{23}c_{12}s_{13}e^{i\delta}&
-s_{23}c_{12}-c_{23}s_{12}s_{13}e^{i\delta}&c_{13}c_{23}
\end{array}\right).
\label{unitmixU1}
\end{eqnarray}
Here $c_{12}=\cos\theta_{12}$,  $s_{12}=\sin\theta_{12}$ etc.

If neutrinos with definite masses are Majorana particles, the mixing matrix  is given by the expression
\begin{equation}\label{Mjmatr}
U^{M}=U^{D}~S,
\end{equation}
where $S=\mathrm{diag}(e^{i\alpha_{1}},e^{i\alpha_{2}},1 )$ is a diagonal phase matrix. From (\ref{mixrel3}) follows that Majorana phases $\alpha_{1,2}$ do not enter into neutrino transition probabilities \cite{Bilenky:1980cx,Doi:1980yb}.

In the Table I we present values of neutrino oscillation parameters obtained from recent global analysis of the  neutrino oscillation data \cite{Gonzalez-Garcia:2014bfa}.
\begin{center}
 Table I
\end{center}
\begin{center}
The values of neutrino oscillation parameters
\end{center}
\begin{center}
\begin{tabular}{|c|c|c|}
  \hline  Parameter &  Normal Spectrum& Inverted Spectrum\\
\hline   $\sin^{2}\theta_{12}$& $0.304^{+0.013}_{-0.012}$& $0.304^{+0.013}_{-0.012}$
\\
\hline    $\sin^{2}\theta_{23}$& $0.452^{+0.052}_{-0.028}$& $ 0.579^{+0.025}_{-0.037}$
\\
\hline   $\sin^{2}\theta_{13}$ & $ 0.0218^{+0.0010}_{-0.0010}$&  $0.0219^{+0.0011}_{-0.0010}$
\\
\hline   $\delta $~(in $^{\circ}$) & $(306^{+39}_{-70})$& $ (254^{+63}_{-62})$
\\
\hline $\Delta m^{2}_{S}$& $(7.50^{+0.19}_{-0.17})\cdot 10^{-5}~\mathrm{eV}^{2}$&$(7.50^{+0.19}_{-0.17})\cdot 10^{-5}~\mathrm{eV}^{2}$\\
\hline $\Delta m^{2}_{A}$& $(2.457^{+0.047}_{-0.047})\cdot 10^{-3}~\mathrm{eV}^{2}$&$(2.449^{+0.048}_{-0.047})\cdot 10^{-3}~\mathrm{eV}^{2}$\\
\hline
\end{tabular}
\end{center}
As we see from this Table, existing neutrino oscillation data do not allow to distinguish normal and inverted neutrino mass spectra. Neutrino oscillation parameters  are known at present with accuracies from about  3 \% ($\Delta m^{2}_{S,A}$)  to about 10\% ($\sin^{2}\theta_{23}$).

Neutrino oscillation data allow to determine only neutrino mass-squared differences. Absolute values of the neutrino masses  at present are unknown. From the measurement of the high-energy part of the $\beta$-spectrum of tritium in Mainz \cite{Kraus:2004zw} and Troitsk \cite{Aseev:2011dq} experiments it was found, respectively,
\begin{equation}\label{MainzTroitsk}
m_{\beta} < 2.3~\mathrm{eV}~(\mathrm{Mainz})~~  m_{\beta}< 2.05~\mathrm{eV} ~(\mathrm{Troitsk})
\end{equation}
Here $m_{\beta}=(\sum_{i}|U_{ei}|^{2}m^{2}_{i})^{1/2}$.

From the recent results of the Planck and other cosmological measurements for the sum of the neutrino masses   it was obtained  the following bound \cite{Ade:2013zuv}
\begin{equation}\label{sumnu}
\sum_{i}m_{i}<0.23
 ~\mathrm{eV}.
\end{equation}
From these bounds it follows that neutrino masses are  much smaller than masses of other fundamental fermions (leptons and quarks). By this reason it is  unlikely that neutrino masses are of the same Standard Model Higgs origin as masses of quarks and leptons. Small neutrino masses are commonly considered as a signature of a beyond the Standard Model physics. However, at present a mechanism of a generation of neutrino masses and neutrino mixing is unknown. In this introductory section we will briefly consider general possibilities for neutrino masses and mixing (see reviews \cite{Bilenky:1987ty,Bilenky:1998dt}).

Masses and mixing are characterized by a  mass term which (in the fermion case we are interested in) is a sum of Lorenz-invariant products of left-handed and right-handed components of the fields. For charged particles only Dirac mass term is allowed. Because electric charges of neutrinos are equal to zero three neutrino mass terms are possible.
The left-handed flavor fields $\nu_{l L}(x)$, which enter into interaction, must enter also into the neutrino mass term. The type of the neutrino mass term depends on the presence in it of right-handed fields $\nu_{l R}(x)$ and on the total lepton number conservation.

\begin{center}
{\bf The Standard Dirac mass term}
\end{center}
If in the Lagrangian there are left-handed and right-handed fields $\nu_{l L}(x) $ and  $\nu_{l R}(x) $ and the total lepton number is conserved the neutrino mass term has the form
\begin{equation}\label{1Dmass}
\mathcal{L}^{\mathrm{D}}(x)=- \sum_{l'l}\bar\nu_{l'
 L}(x)\,
M_{l'l}^{\mathrm{D}}\,\nu_{l R}(x)
+\rm{h.c.}
\end{equation}
A complex $3\times3 $ matrix  $M^{\mathrm{D}}$ can be presented in the form
\begin{equation}\label{1Dmass1}
M^{\mathrm{D}}=U~m~V^{\dag},
\end{equation}
where $U$ and $V$ are unitary mixing matrix and $m$ is a diagonal matrix. From (\ref{1Dmass}) and (\ref{1Dmass1}) we find
\begin{equation}\label{1Dmass2}
\mathcal{L}^{\mathrm{D}}(x)=-
   \sum_{i=1}^{3}  m_{i}\,\bar\nu_{i}(x)\,\nu_{i}(x)
\end{equation}
Thus, $\nu_{i}$ is the field of  neutrino with mass $m_{i}$.

The flavor fields $\nu_{l L}(x)$ are connected with the fields
$\nu_{iL}(x)$ by the  mixing relation
\begin{equation}\label{1Dmass3}
\nu_{l L}(x)=\,\sum^{3}_{i=1}U_{l i}\,\nu_{i L}(x),
\end{equation}
where $U$ is  the unitary PMNS  mixing matrix which is characterized  by three mixing angles and one $CP$ phase.

In the case of the mass term (\ref{1Dmass}) the invariance under the global gauge transformations
\begin{equation}\label{1Dmass4}
    \nu'_{lL}(x)=e^{i\alpha}~\nu_{l L}(x),~~~ \nu'_{lR}(x)=e^{i\alpha}~\nu_{lR}(x) \quad  l'_{L,R}(x)=e^{i\alpha} ~l_{ L,R}(x)\quad l=e,\mu,\tau
\end{equation}
takes place ($\alpha$ is a constant phase, same for all flavors). The invariance under the transformations (\ref{1Dmass4}) means that the total lepton number $L$
is conserved and $\nu_{i}(x)$  is the Dirac field of neutrinos ($L(\nu_{i})=1$) and antineutrinos ($L(\bar\nu_{i})=-1$). The mass term (\ref{1Dmass}) is the standard Dirac mass term.

\begin{center}
{\bf The most economical Majorana mass term}
\end{center}
If there are only left-handed fields $\nu_{l L}(x)$ in the Lagrangian we can  built the neutrino mass term if we take into account that $(\nu_{l L})^{c}=C \bar\nu_{l L}^{T}$ is a right-handed component ($C$ is the matrix of the charge conjugation which satisfies the relations $C\gamma^{T}_{\alpha}C^{-1}=-\gamma_{\alpha},~~C^{T}=-C$). For the mass term we have in this case
\begin{equation}\label{1Mjmass}
\mathcal{L}^{\mathrm{L}}(x)= -\frac{1}{2}\,\sum_{l',l}
\bar\nu_{l'L}(x)\,M_{l'l}^{L} ( \nu_{lL})^{c}(x)
+\mathrm{h.c.}
\end{equation}
Here $M^{L}$ is a complex symmetrical $3\times 3$ matrix. The matrix $M^{L}$ can be presented in the form
\begin{equation}\label{1Mjmass1}
M^{L}=U~m~U^{T},
\end{equation}
where $U$ is a unitary matrix, and $m_{ik}=m_{i}~\delta_{ik},~m_{i}>0$. From (\ref{1Mjmass}) and (\ref{1Mjmass1}) we have
\begin{equation}\label{1Mjmass2}
\mathcal{L}^{\mathrm{L}}(x)=-\frac{1}{2}\sum_{i=1}^{3}  m_{i}\,\bar\nu_{i}(x)\,\nu_{i}(x),
\end{equation}
where the field  $\nu_{i}(x)$ (i=1,2,3) satisfies the condition
\begin{equation}\label{1Mjmass3}
\nu_{i}(x)=\nu^{c}_{i}(x)=C\bar\nu_{i}^{T}(x).
\end{equation}
Thus $\nu_{i}(x)$ is the field of the truly neutral

 Majorana neutrino ( $\nu_{i}\equiv \bar \nu_{i}$)  with mass $m_{i}$. The flavor field $\nu_{l L}(x)$ is connected with
left-handed components $\nu_{i L}$  by the  mixing relation
\begin{equation}\label{1Mjmass4}
\nu_{l L}(x)=\,\sum^{3}_{i=1}U_{l i}\,\nu_{i L}(x)~~l=e,\mu,\tau,
\end{equation}
 The unitary mixing matrix $U$  is characterized  by three mixing angles and three $CP$ phases.

The Lagrangian (\ref{1Mjmass}) is not invariant under the global gauge transformation $\nu'_{l L}(x)=e^{i\alpha}\nu_{l L}(x)$. Thus, in the case of the mass term (\ref{1Mjmass})  the total lepton number $L$ is not conserved and there is no conserved the quantum number which can distinguish neutrino and antineutrino. This is the reason why fields of neutrinos with definite masses $\nu_{i}(x)$ are Majorana fields.

\begin{center}
{\bf The most general Dirac and Majorana mass term}
\end{center}
The most general neutrino mass term has the form
\begin{equation}\label{1DMmass}
\mathcal{L}^{\mathrm{D+M}}(x)=\mathcal{L}^{\mathrm{L}}(x)+\mathcal{L}^{\mathrm{D}}(x)+\mathcal{L}^{\mathrm{R}}(x)
\end{equation}
Here
\begin{equation}\label{1DMmasss}
\mathcal{L}^{\mathrm{R}}(x)=-\frac{1}{2}\,\sum_{l'l}\overline{(\nu_{l'R}(x))^{c}}\,M_{l'l}^{R} \nu_{lR}(x)
+\mathrm{h.c.}
\end{equation}
and $\mathcal{L}^{\mathrm{D}}(x)$ and $\mathcal{L}^{\mathrm{L}}(x)$   are given, respectively, by (\ref{1Dmass}) and (\ref{1Mjmass}).

The mass term (\ref{1DMmass}) does not conserve the total lepton number $L$. After the diagonalization of the mass term
 we have
\begin{equation}\label{1DMmass1}
\mathcal{L}^{\mathrm{D+M}}(x)=-\frac{1}{2}\sum_{i=1}^{6}  m_{i}\,\bar\nu_{i}(x)\,\nu_{i}(x)
\end{equation}
where $\nu_{i}(x)$   is the Majorana field  with mass $m_{i}$:
\begin{equation}\label{1DMmass2}
    \nu_{i}(x)=\nu^{c}_{i}(x)=C\bar\nu_{i}(x)^{T}\quad i=1,2...6.
\end{equation}
The flavor fields $\nu_{l L}(x)$ and the fields $(\nu_{l R}(x))^{c}$ are connected with the left-handed components of the Majorana fields $\nu_{i L}(x)$ by the following mixing relations
\begin{equation}\label{1DMmass3}
\nu_{l L}(x)=\,\sum^{6}_{i=1}U_{l i}\,\nu_{i L}(x),\quad(\nu_{l R}(x))^{c}=
\,\sum^{6}_{i=1}U_{\bar {l} i}\,\nu_{i L}(x)~~l=e,\mu,\tau.
\end{equation}
Here $U$ is a unitary $6\times 6$ mixing matrix.

Right-handed neutrino fields $\nu_{l R}(x)$ do not enter into
the SM Lagrangian and are called sterile fields. As we see from (\ref{1DMmass3}), in the case of the Dirac and Majorana mass term the flavor fields $\nu_{l L}$ are mixtures of six left-handed components of  Majorana fields $\nu_{i L}$.
Sterile fields $(\nu_{l R}(x))^{c}$ are mixtures of the same six left-handed components.

Different possibilities can be considered in the case of  the Dirac and Majorana mass term.
The most popular are the following.
\begin{enumerate}
  \item Transitions into sterile states.

If the number of light Majorana neutrinos $\nu_{i}(x)$ is larger  than three,  transitions of flavor neutrinos into sterile neutrinos become possible. For the neutrino mixing we have in this case
\begin{equation}\label{1DMmass4}
\nu_{\alpha L}(x)=\,\sum^{3+n_{s}}_{i=1}U_{\alpha i}\,\nu_{i L}(x),\quad  \alpha=e,\mu,\tau, s_{1},...
\end{equation}
where $n_{s}$ is the number of the sterile neutrinos.

There exist at present some indications in favor of transitions of flavor neutrinos into sterile states. We will discuss these indications later.

\item  Seesaw mechanism of the neutrino mass generation.

If in the  spectrum of masses of the Majorana particles there are three light (neutrino) masses and three  heavy masses, we can explain smallness of neutrino masses with respect to the masses of leptons and quarks. This is the  famous seesaw mechanism of the neutrino mass generation \cite{Minkowski:1977sc,Yanagida:1979as,GellMann:1980vs,Glashow:1979nm,Mohapatra:1980ia}.  We will consider this mechanism later.

\end{enumerate}
The Dirac mass term can be generated by the Standard Higgs mechanism of the mass generation.  This mechanism can not explain, however, the smallness of neutrino masses. The Majorana mass term and the Dirac and Majorana mass term can be generated only by beyond the SM mechanisms. At the moment we do not know the type of neutrino mixing:
all possibilities are  open. Later we will discuss the most plausible and economical possibility.

\section{On the Standard Model of the electroweak interaction}

\subsection{Introduction}
The Standard Model \cite{Glashow:1961tr,Weinberg:1967tq,Salam:1968rm}  is one of the greatest achievement of the physics of the XX's century. It emerged as a result of numerous experiments and  fundamental theoretical principles (local gauge invariance and others). After discovery of the Higgs boson at LHC the Standard Model got the status of {\em the theory} of physical phenomena in the electroweak energy scale (up to about 300 GeV).  We will try here to make  some general conclusions which can be inferred from the Standard Model and   apply them to neutrinos.

There are many questions connected with the Standard Model: why left-handed and right-handed quark, lepton and neutrino fields have different transformation properties, why in unified electroweak interaction the weak CC part maximally violate parity and the electromagnetic part conserve parity etc. I suggest here that the CC structure of the Standard Model and such her features are  due to neutrinos.

The Standard Model is based on the following principles
\begin{enumerate}
  \item Local gauge symmetry.
  \item Unification of the  electromagnetic and weak interactions into one electroweak interaction.
  \item Spontaneous breaking of the electroweak symmetry.
\end{enumerate}
We will demonstrate here that in the framework of these principles nature choose the simplest, most economical possibilities.

\subsection{Two-component neutrino}

 {\em From my point of view the SM started with the theory of the two-component neutrino.} First of all some historical remark.

In 1929 soon after Dirac proposed his famous equation for four-component spinors, which describe relativistic particle with spin 1/2, Weyl published a paper \cite{Weyl}  in which he introduced two-component spinors. For a particle with spin 1/2 Weyl wanted  to built equation for the two-component wave function, like the Pauli one, but  Lorenz-invariant.
He came to a conclusion that this is impossible if mass of the particle is not equal to zero. For a massless particle he found equations
\begin{equation}\label{weyl}
i\gamma^{\alpha}\partial_{\alpha}~\psi_{L}(x)=0,\quad i\gamma^{\alpha}\partial_{\alpha}~\psi_{R}(x)=0,
\end{equation}
where $\psi_{L}(x)$ and $\psi_{R}(x)$ are left-handed and right-handed two-component spinors which satisfy the conditions
\begin{equation}\label{weil}
\gamma _{5}\psi_{L,R}(x)=\mp \psi_{L,R}(x).
\end{equation}
Under the inversion of the coordinates the left-handed (right-handed) spinor is transformed into right-handed (left-handed) spinor:
\begin{equation}\label{weil1}
\psi'_{R,L}(x')=\eta \gamma^{0}\psi_{L,R}(x),\quad x'=(x^{0},-\vec{x}).
\end{equation}
Here $\eta$ is a phase factor. Thus, Weyl equations (\ref{weyl}) are not invariant under the inversion (do not conserve parity).

At the time when Weyl proposed the equations (\ref{weyl}) (and many years later) physicists believed that the conservation of the parity is the law of the nature. So, Weyl theory was rejected.\footnote{Pauli in his book on Quantum Mechanics \cite{Pauli} wrote "...because the equation for
$\psi_{L}(x)$ ($\psi_{R}(x))$ is not invariant under space reflection it is not applicable to the physical reality".
Notice, however, the following statement which belong to Weyl "My work  always tried to unite the truth with the beautiful, but when I had to choose one or the other, I usually choose the beautiful."}

After  it was discovered \cite{Wu,Lederman} (1957)  that parity is not conserved in the $\beta$-decay and other weak processes, Landau \cite{Landau:1957tp}, Lee and Yang \cite{Lee:1957qr} and Salam \cite{Salam:1957st}  proposed the theory of the two-component neutrino. These authors had different arguments in favor of such a theory. Landau built CP-invariant neutrino theory, Salam considered $\gamma_{5}$ invariant theory and Lee and Yang applied to neutrino the Weyl's theory.

The authors of the two-component neutrino theory  assumed that neutrino mass is equal to zero (which was compatible with existed at that time data) and that neutrino field was $\nu_{L}(x)$ or
$\nu_{R}(x)$. Such fields  satisfy the Weyl equations
\begin{equation}\label{weyl1}
i\gamma^{\alpha}\partial_{\alpha}~\nu_{L}(x)=0, ~~~i\gamma^{\alpha}\partial_{\alpha}~\nu_{R}(x)=0.
\end{equation}
If neutrino is the two-component  particle in this case
\begin{enumerate}
  \item Large violation of the parity in the $\beta$-decay, $\mu$-decay and other weak processes must be observed (in agreement with the results of the  Wu et al and other experiments \cite{Wu:1957my,Lederman}).

  \item Neutrino (antineutrino) helicity is equal to -1 (+1) in the case of the field $\nu_{L}(x)$ and is equal to +1 (-1) in the case of the field $\nu_{R}(x)$.
\end{enumerate}
The point 1. is obvious from (\ref{weil1}). In order to see that two-component neutrino is a particle with definite helicity  let us consider the spinor $u^{r}(p)$ which describes a massless particle with the momentum $p$ and helicity $r$. We have
\begin{equation}\label{helicity}
\gamma\cdot p ~ u^{r}(p)=0,\quad \vec{\Sigma}\cdot \vec{n}~u^{r}(p)=r~u^{r}(p),~~~r=\pm 1.
\end{equation}
Here $\vec{\Sigma}=\gamma_{5}\gamma^{0}\vec{\gamma}$  is the operator of the spin and $\vec{n}=\frac{\vec{p}}{|\vec{p}|}$ is the unit vector in the direction of the neutrino momentum. From (\ref{helicity}) it follows that
\begin{equation}\label{helicity1}
 \gamma_{5}~u^{r}(p)=r~u^{r}(p).
\end{equation}
 Thus, we have
\begin{equation}\label{1helicity}
 \frac{1}{2}(1\mp \gamma_{5})~u^{r}(p)=\frac{1}{2}(1\mp r)~u^{r}(p)
\end{equation}
From this relation it follows that  $r=-1~ (r=+1)$ if neutrino field is $\nu_{L}(x)$ ($\nu_{R}(x)$).
 Analogously, it is easy to show that antineutrino helicity is equal to +1 (-1) in the case if neutrino field is
$\nu_{L}(x)$ ($\nu_{R}(x)$).

 The neutrino helicity was measured in the spectacular
Goldhaber et al experiment \cite{Goldhaber:1958nb}.  In this experiment the neutrino helicity was obtained from the measurement of the circular polarization of $\gamma$-quanta produced in the chain of reactions
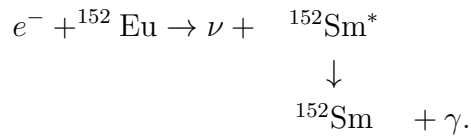
\begin{eqnarray}\label{helicity2}
e^- + ^{152}\rm {Eu} \to \nu + \null & ^{152}\rm {Sm}^* & \nonumber
\\
& \downarrow & \nonumber
\\
& ^{152}\rm{Sm} & \null + \gamma.\nonumber
\end{eqnarray}
The authors of the paper  \cite{Goldhaber:1958nb} concluded : "... our result is compatible with 100\% negative helicity  of neutrino emitted in orbital electron capture".

Thus, the Goldhaber et al experiment confirmed the two-component neutrino theory. It was shown that from two possibilities ($\nu_{L}(x)$ or $\nu_{R}(x)$) nature choose  the first one.

Let us notice that at the time when  the two-component neutrino theory was proposed it was unknown that exist three  types of neutrino. In 1962 in the Brookhaven experiment \cite{Danby:1962nd} it was shown that muon and electron neutrinos $\nu_{e}$ and $\nu_{\mu}$ are different particles. In 2000 the third neutrino $\nu_{\tau}$ was discovered in the DONUT experiment \cite{Kodama:2000mp}.

The number of degrees of freedom of the two-component Weyl field is two times smaller than the number of the degrees of freedom of the four-component Dirac field. It looks plausible that  {\em for neutrino nature chooses this simplest and most economical possibility.}

\subsection{Local gauge symmetry}

The local gauge symmetry is a natural symmetry for the Quantum Field Theory with quantum fields which depend on $x$. In accordance with the two-component neutrino theory we will assume that the fields of electron, muon and tau neutrinos are left-handed  two-component Weyl fields. We will denote them  $\nu'_{eL}, \nu'_{\mu L}, \nu'_{\tau L}$. Neutrinos
$\nu_{e}, \nu_{\mu}, \nu_{\tau}$ take part in the CC weak interaction together with, correspondingly, $e, \mu, \tau$. The requirements of the symmetry can be satisfied if  electron, muon and tau fields, like neutrino fields, are also left-handed two-component Weyl fields ($e'_L, \mu'_L, \tau'_L$). The simplest symmetry group is $SU_{L}(2)$ and the simplest possibility for neutrino and lepton fields is to be, correspondingly, up and down components of the doublets:\footnote{We will consider only leptons. Notice also that meaning of primes will be clear later.}
\begin{eqnarray}
\psi^{lep}_{eL}=\left(
\begin{array}{c}
\nu'_{eL} \\
e'_L \\
\end{array}
\right),~\psi^{lep}_{\mu L}=\left(
\begin{array}{c}
\nu'_{\mu L} \\
\mu'_L \\
\end{array}
\right),~\psi^{lep}_{\tau
 L}=\left(
\begin{array}{c}
\nu'_{\tau L} \\
\tau'_L \\
\end{array}
\right).\label{lepSU2dub}
\end{eqnarray}
In order to insure the invariance under the local gauge transformations
\begin{equation}\label{locSU2}
(\psi^{lep}_{l})'(x) = e^{i\,\frac{1}{2}~\vec\tau\cdot\vec\Lambda(x)}~\psi^{lep}_{l}(x)\quad (l=e,\mu,\tau)
\end{equation}
($\vec\tau\cdot\vec\Lambda(x)=\sum^{3}_{i=1}\tau^{i}\Lambda^{i}(x)$, $\tau^{i}$ are Pauli matrices and                  $\Lambda^{i}(x)$ are {\em arbitrary  functions of $x$}) we need to assume that neutrino-lepton  fields interact with massless vector  fields $\vec A_{\alpha}(x)$ and in the free Lagrangian derivatives of the fermion fields are changed by the covariant derivatives
\begin{equation}\label{1covder}
\partial_{\alpha} \,\psi^{lep}_{lL}(x)\to  (\partial_{\alpha} +
i\,g\,\frac{1}{2}\,\vec\tau\cdot\vec A_{\alpha}(x))\,\psi^{lep}_{lL}(x),
\end{equation}
where $g$ is a dimensionless  constant and the field $\vec A_{\alpha}(x)$ is transferred as follows
\begin{equation}\label{2covder}
\vec A'_{\alpha}(x)=\vec A_{\alpha}(x)-\frac{1}{g}\partial_{\alpha}\vec \Lambda(x)-\vec \Lambda(x)\times \vec A_{\alpha}(x).
\end{equation}
With the change (\ref{1covder}) we generate the following Lagrangian of the interaction of the lepton and vector  $\vec A_{\alpha}(x)$ fields
\begin{equation}\label{interL}
\mathcal{L}_{I}(x) = -g~\vec{j}_{\alpha}(x)\vec{A}^{\alpha}(x).
\end{equation}
Here
\begin{equation}\label{isocurrent}
\vec{j}_{\alpha}=\sum_{l=e,\mu,\tau}\bar\psi^{lep}_{lL}\gamma_{\alpha}\frac{1}{2}\vec{\tau} \psi^{lep}_{lL}
\end{equation}
is the isovector  current.

The expression (\ref{interL}) can be written in the form
\begin{equation}\label{interL1}
\mathcal{L}_{I}(x) = \left( -\frac{g}{2\,\sqrt{2}}\,
j^{CC}_{\alpha}(x)\,W^{\alpha}(x) + \rm{h.c}\right)
-g\,j^{3}_{\alpha}(x)\,A^{3\alpha }(x)~.
\end{equation}
Here
\begin{equation}\label{3CCcurrent}
j^{CC}_{\alpha}=2(j^{1}_{\alpha}+ij^{2}_{\alpha}) = 2\sum_{l=e,\mu,\tau}\bar\nu'_{lL}\gamma_{\alpha} l'_L
\end{equation}
is the lepton charged current and $W_{\alpha}=\frac{A^{1}_{\alpha}-iA^{2}_{\alpha}}{\sqrt{2}}$ is the field of charged, vector $W^{\pm}$ bosons.

The following remarks are in order:
\begin{enumerate}
  \item The local gauge invariance {\em requires} existence of the vector field $\vec{A}^{\alpha}(x)$. This field is called gauge vector field.
  \item The interaction (\ref{interL}) is {\em the minimal interaction} compatible with local gauge invariance.
  \item From (\ref{2covder}) it follows that the strength tensor of the vector field $\vec{A}^{\alpha}(x)$ is given by the expression
\begin{equation}\label{stresstenz}
\vec F_{\alpha \beta}(x)= \partial_{\alpha}\,\vec A_{\beta}(x)
-\partial_{\beta}\,\vec A_{\alpha}(x)-g\,\vec
A_{\alpha}(x)\times\vec A_{\beta}(x),
\end{equation}
where the last term is due to the fact that $SU(2)_{L}$ is a non-Abelian group. Because interaction constant $g$ enters into expression for the strength tensor it must be the same for all doublets $\psi^{lep}_{lL}(x)$ ($l=e,\mu,\tau$).
As a result we came to {\em $e-\mu-\tau$ universal charged current weak interaction (\ref{interL1}}).
\end{enumerate}

\subsection{Unification of the weak and electromagnetic interactions}

{\em The Standard Model is the unified theory of the weak and electromagnetic interactions}. In the electromagnetic current of the charged leptons enter left-handed and right-handed fields :
\begin{equation}\label{EMcur}
j^{\mathrm{EM}} _{\alpha}= \sum_{l}(-1)~\bar l'\gamma_{\alpha} l'=\sum_{l}(-1)~\bar l_{L}'\gamma_{\alpha} l'_{L}
+\sum_{l}(-1)~\bar l_{R}'\gamma_{\alpha} l'_{R}.
\end{equation}
Thus, in order to unify  weak and electromagnetic interactions we must enlarge symmetry group. A new symmetry group must include not only transformations of left-handed fields but also transformations of right-handed fields {\em of charged leptons.} There is a fundamental difference between neutrinos and other  fermions: neutrinos electric charges are equal to zero, there is no electromagnetic current of neutrinos. The   unification of the weak and electromagnetic interactions  does not  require right-handed neutrino fields.  {\em A minimal possibility is to assume that  there are no right-handed neutrino fields in the Standard Model.}

The minimal enlargement of the $SU_{L}(2)$ group is a direct product $SU_{L}(2)\times U_{Y}(1)$. In order to ensure local gauge $SU_{L}(2)\times U_{Y}(1)$ invariance we need to change in the free Lagrangian derivatives of left-handed and right-handed fields by the covariant derivatives
\begin{eqnarray}\label{covderiv}
\partial_{\alpha}\psi^{lep}_{lL} &\to &(\partial_{\alpha} +ig\frac{1}{2}\vec{\tau}\cdot\vec{A_{\alpha}}+ig'\frac{1}{2}Y^{lep}_{L}B_{\alpha})\psi^{lep}_{lL},\nonumber\\
\partial_{\alpha}l'_{R} &\to& (\partial_{\alpha} +ig'\frac{1}{2}Y^{lep}_{R}B_{\alpha})l'_{R},
\end{eqnarray}
where  $B_{\alpha}$ is vector gauge field of the $U_{Y}(1)$ group.

There are no constraints on the interaction constants  of the Abelian $U_{Y}(1)$ local group. In order to unify the weak and electromagnetic interactions we  assume that the interaction constants for lepton doublets  and charged lepton singlets have the form
 \begin{equation}\label{interconst}
 g'\frac{1}{2}Y^{lep}_{L}, ~~~ g'\frac{1}{2}Y^{lep}_{R}.
 \end{equation}
Here $g'$ is a constant and hypercharges of left-handed and right-handed fields $Y^{lep}_{L}$ and $Y^{lep}_{R}$ are determined by the Gell-Mann-Nishijima relation
\begin{equation}\label{3GM-N}
    Q=T_{3}+\frac{1}{2}Y,
\end{equation}
where $Q$ is the electric charge  and $T_{3}$ is the third projection of the isotopic spin.

For the Lagrangian of the minimal interaction of the lepton  fields and the fields $A^{3}_{\alpha}$ and $B_{\alpha}$ of  neutral vector bosons we obtain the following expression
\begin{equation}\label{interL5}
\mathcal{L}^{0}_{I}=-g\,j^{3}_{\alpha}\,A^{3\alpha }-g'~\frac{1}{2}~j^{Y} _{\alpha}~B^{\alpha}.
\end{equation}
Here
 \begin{equation}\label{interL6}
 \frac{1}{2}~j^{Y} _{\alpha}=j^{EM}_{\alpha}-j^{3}_{\alpha},
 \end{equation}
where $j^{EM}_{\alpha}$ is the electromagnetic current of the leptons.

Notice that the electromagnetic current appeared in (\ref{interL6}) due to the fact that electric charges of  left-handed components $l'_{L}$ (coming from doublets) and  right-handed components  $l'_{R}$ (coming from singlets) are the same. Thus, if we  choose   coupling constants
of the $U_{Y}(1)$ local gauge group in accordance with the Gell-Mann-Nishijima relation we can combine electromagnetic interaction which conserve parity with the weak interaction which violate parity into one electroweak interaction.

In order to separate in (\ref{interL5}) the Lagrangian of electromagnetic
 interaction of leptons  with the electromagnetic field
\begin{itemize}
  \item  instead of the fields $A^{3 \alpha}$ and $B^{\alpha}$  we introduce  "mixed" fields
 \begin{equation}\label{mixfields}
Z^{ \alpha}=\cos\theta_{W} A^{3 \alpha}-\sin\theta_{W} B^{ \alpha},\quad A^{ \alpha}=\sin\theta_{W} A^{3 \alpha}+
\cos\theta_{W}B^{ \alpha},
\end{equation}
where angle $\theta_{W}$ is determined by the relation
\begin{equation}\label{Wangle}
\frac{g'}{g}=\tan \theta_{W}.
\end{equation}
\item we assume that the following relation holds
\begin{equation}\label{unif}
    g~\sin \theta_{W}=e.
 \end{equation}
Here e is the proton charge. The relation (\ref{unif}) is called the  {\em unification condition}.
\end{itemize}
Finally, the interaction Lagrangian takes the form
\begin{equation}\label{Unif}
\mathcal{L}_{I}=\mathcal{L}^{CC}_{I} +   \mathcal{L}^{NC}_{I}+\mathcal{L}^{EM}_{I}.
\end{equation}
Here
\begin{equation}\label{CCinter}
\mathcal{L}^{CC}_{I} = \left( -\frac{g}{2\,\sqrt{2}}\,
j^{CC}_{\alpha}\,W^{\alpha} + \rm{h.c}\right),
\end{equation}
is  the charged current Lagrangian,
\begin{equation}\label{NCinter}
\mathcal{L}^{NC}_{I}=-\frac{g}{2\cos\theta_{W}}\,j^{\rm{NC}}_{\alpha}\,Z^ {\alpha}.
\end{equation}
is the neutral current Lagrangian,
\begin{equation}\label{EMinter}
\mathcal{L}^{EM}_{I} =
-e~j^{EM}_{\alpha}\,A^{\alpha}
\end{equation}
is the electromagnetic  Lagrangian.

We  considered up to now only neutrinos and charged leptons. If we include also quarks
the total charged,  neutral  and electromagnetic  currents  are given by the following expressions
\begin{equation}\label{1CC1}
j^{CC}_{\alpha}=2\sum_{l=e,\mu,\tau}\bar\nu'_{lL}\gamma_{\alpha} l'_{L}+2(\bar u'_{L}\gamma_{\alpha} d'_{L} +\bar c'_{L}\gamma_{\alpha} s'_{L}+\bar t'_{L}\gamma_{\alpha} b'_{L}),
\end{equation}
\begin{equation}\label{NC}
j^{\rm{NC}}_{\alpha}=2~ j^{3}_{\alpha} -2~\sin^{2}\theta_{W}\,j^{\rm{EM}}_{\alpha},
\end{equation}
where
\begin{equation}\label{1NC1}
 j^{3}_{\alpha}=\frac{1}{2}\sum_{l=e,\mu,\tau}\bar\nu'_{lL}\gamma_{\alpha} \nu'_{lL}-\frac{1}{2}\sum_{l=e,\mu,\tau}\bar l'_{L}\gamma_{\alpha} l'_{L}+\frac{1}{2}\sum_{q=u,c,t}\bar q'_{L}\gamma_{\alpha}q'_{L}
-\frac{1}{2}\sum_{q=d,s,b}\bar q'_{L}\gamma_{\alpha}q'_{L}.
\end{equation}
and
\begin{equation}\label{1EM}
j^{EM}_{\alpha}=(-1)\sum_{l=e,\mu,\tau}\bar l'\gamma_{\alpha}l'+(\frac{2}{3})\sum_{q=u,c,t}\bar q'\gamma_{\alpha}q'+
(\frac{-1}{3})\sum_{q=d,s,b}\bar q'\gamma_{\alpha}q'.
\end{equation}
Let us stress that the structure of the CC term is determined the two-component neutrinos. The structure of the NC term is determined by the unification the CC and EM interactions on the basis of  the $SU_{L}(2)\times U_{Y}(1)$ group.

The  Lagrangian of interaction of fundamental fermions and gauge vector bosons {\em is the minimal, simplest Lagrangian}. However, due to requirements of the local gauge $SU_{L}(2)\times U_{Y}(1)$  symmetry there are no mass terms of all fermions and gauge vector bosons in the  Lagrangian.

In order to build a realistic theory of the electroweak interaction we need to violate local gauge symmetry and generate
masses of $W^{\pm}$ and $Z^{0}$ bosons and mass terms of quarks and charged leptons. The photon must remain massless. Neutrino masses is a special subject. We will discuss it later.

\subsection{Brout-Englert-Higgs spontaneous symmetry breaking}

{\em The Standard model mechanism of the mass generation is  the Brout-Englert-Higgs mechanism} \cite{Englert:1964et,Higgs:1964ia,Higgs:1964pj}. It is based on the phenomenon of the spontaneous symmetry breaking. The spontaneous symmetry breaking takes place in the ferromagnetism and other  many-body phenomena. It happens if the Hamiltonian of the system has some symmetry and vacuum states are degenerated.   It was suggested  \cite{Nambu:1960xd,Nambu:1961fr,Goldstone:1961eq} that the phenomenon of the spontaneous symmetry breaking takes place also in the Quantum Field Theory.

In order to ensure the spontaneous symmetry breaking  in addition to the fields of fundamental fermions and gauge vector bosons {\em we must include also the scalar Higgs field in the system.}

We will assume that the Higgs field
\begin{equation}\label{Hdoub}
\phi(x) ={\phi _{+}(x)\choose\phi _{0}(x)}
\end{equation}
is transformed as  $SU_{L}(2)$ doublet. Here $\phi _{+}(x)$ and $\phi _{0}(x) $ are  complex  charged and neutral scalar fields. According to  the Gell-Mann-Nishijima relation the hypercharge of the doublet $\phi(x)$ is equal to one.
We will see later that {\em this assumption give us the  most economical possibility to generate masses of $W^{\pm}$ and $Z^{0}$ vector bosons.}

The  part of  $SU_{L}(2)\times U_{Y}(1)$ invariant Lagrangian, in which the Higgs field enters,  has the form
\begin{equation}\label{HiggsL}
 \mathcal{L}=
((\partial_{\alpha}+i\,g\,\frac{1}{2}\,\vec\tau\cdot\vec
A_{\alpha} + i\,g'\,\frac{1}{2}\, B_{\alpha} )\,\phi
)^{\dagger}
(\partial^{\alpha}+i\,g\,\frac{1}{2}\,\vec\tau\cdot\vec
A^{\alpha} + i\,g'\,\frac{1}{2}\,
B^{\alpha} \,)\,\phi
-V(\phi^{\dagger}\,\phi),
\end{equation}
where potential $V(\phi^{\dagger}\,\phi)$  is given by the expression
\begin{equation}\label{Higgspot}
V(\phi^{\dagger}\,\phi)= -\mu^{2}\,\phi^{\dagger}\,\phi
+\lambda\,(\phi^{\dagger}\,\phi)^{2}.
\end{equation}
Here $\mu^{2}$ and $\lambda$ are positive constants. The constant $\mu$ has dimension $M$ and the constant $\lambda$ is dimensionless constant.

Existence of the Higgs field fundamentally change properties of the system:  the energy of the system reaches  minimum {\em at nonzero values of the Higgs field}. In fact, the  energy reaches the minimum at such values of Higgs field which minimize the potential. We can rewrite the potential in the form
\begin{equation}\label{Higgspot1}
V(\phi^{\dagger}\,\phi)=\lambda~\left(\phi^{\dagger}\,\phi-
\frac{\mu^{2}}{2\lambda}\right)^{2}-\frac{\mu^{4}}{4\lambda}.
\end{equation}
From this expression it is obvious that the potential reaches minimum at
\begin{equation}\label{Hmin}
(\phi^{\dagger}\,\phi)_{0}= \frac{v^{2}}{2}
\end{equation}
where
\begin{equation}\label{Hmin1}
v^{2}=\frac{\mu^{2}}{\lambda}.
\end{equation}
Taking into account the conservation of the electric charge, for the vacuum values of the Higgs field we have
\begin{equation}\label{vev5}
    \phi_{0} = {0\choose\frac {v}{\sqrt{2}} }~e^{i\alpha},
\end{equation}
where $\alpha$ is an arbitrary phase. It is obvious that this freedom is due to the gauge symmetry of the Lagrangian.
We can choose
\begin{equation}\label{vev6}
    \phi_{0} = {0\choose\frac {v}{\sqrt{2}} }.
\end{equation}
With this choice we  break the symmetry. Notice that in the quantum case the constant $v$, having the dimension $M$,
is the vacuum expectation value (vev) of the Higgs field.

The doublet $\phi (x)$ can  be presented in the form
\begin{equation}
\begin{array}{l}
\phi(x) = e^{i\,\frac{1}{v}\frac{1}{2}\,\vec\tau\cdot\vec\theta (x)}~
\left(\begin{array}{c} 0\\\frac{v+H(x)}{\sqrt{2}}
\end{array}\right).
\end{array}
\label{Hpresent}
\end{equation}
Here  $\theta_{i} (x)$ ($i=1,2,3)$ and $H(x)$ are real functions which have dimension of the scalar field ($M$). Vacuum values of these functions are equal to zero.

The Lagrangian (\ref{HiggsL}) is invariant under  $SU_{L}(2)\times U_{Y}(1)$ local gauge transformations. We can choose the arbitrary gauge in such a way that
\begin{equation}
\begin{array}{l}
\phi(x) =
\left(\begin{array}{c} 0\\\frac{v+H(x)}{\sqrt{2}}
\end{array}\right).
\end{array}
\label{Higgsfield}
\end{equation}
Such a gauge is called the unitary gauge. From (\ref{Higgsfield}) it follows that the Lagrangian (\ref{HiggsL}) takes the form
\begin{equation}\label{hL4}
\mathcal{L}=\frac{1}{2} ~\partial_{\alpha}H\,\partial^{\alpha}H
+\frac{1}{4}\, (v +H)^{2}g^{2} W^{\dag}_{\alpha}\,W^{\alpha}+
\frac{1}{4}\,(v +H)^{2}(g^{2}+g'^{2})\frac{1}{2}Z_{\alpha}\,Z^{\alpha}-
\frac{\lambda}{4}\, (2v H +H^{2})^{2}.
\end{equation}
The mass terms of $W^{\pm}$ and $Z^{0}$ vector bosons and  the scalar Higgs boson are given by the expressions
\begin{equation}\label{massterm}
\mathcal{L}^{m}=m^{2}_{W}\,W^{\dag}_{\alpha}\,W^{\alpha} +
\frac{1}{2}\,m^{2}_{Z}\,Z_{\alpha}\,Z^{\alpha} -
\frac{1}{2}\,m^{2}_{H}\,H^{2},
\end{equation}
where $m_{W}$, $m_{Z}$ and $m_{H}$ are  masses of  $W^{\pm}$, $Z^{0}$ and  Higgs bosons.
From (\ref{hL4}) and (\ref{massterm}) we find
\begin{equation}\label{masses}
m_{W} =\frac{1}{2}\, g\, v ,\quad
m_{Z}=\frac{1}{2}\sqrt{(g^{2}+g^{'2})}~ v,\quad m_{H}=
\sqrt{2\lambda}~v .
\end{equation}
Thus, after the spontaneous symmetry breaking,  $W^{\alpha}(x)$ becomes
the field of the charged vector $W^{\pm}$ bosons with the mass
$\frac{1}{2}gv$, $Z^{\alpha}(x)$ is the field of neutral vector $Z^{0}$
bosons with the mass $\frac{1}{2}\sqrt{(g^{2}+g^{'2})
}~v$, $A_{\alpha}(x)$ remains the  field of massless photons.

 Three (Goldston) degrees of freedom are necessary to
provide longitudinal components of massive $W^{\pm}$ and $Z^{0}$ bosons. The Higgs doublet (two complex scalar fields, 4 degrees of freedom) is a {\em minimal possibility.} One remaining degree of freedom is a neutral  Higgs field $H(x)$
of scalar particles with the mass $\sqrt{2\lambda}~v$.

{\em The  Brout-Englert-Higgs mechanism of the generation of masses of  $W^{\pm}$ and $Z^{0}$ bosons  predicts  existence of the massive scalar boson.} Recent discovery of the scalar boson at LHC \cite{Aad:2012tfa,Chatrchyan:2012ufa} is an impressive confirmation of this prediction of the Standard Model.

The expressions (\ref{masses}) for masses of the $W^{\pm}$ and $Z^{0}$ bosons are characteristic expressions for masses of  vector bosons in a theory   with spontaneous symmetry breaking (and  covariant derivative of the Higgs field in the Lagrangian). In fact, it is evident from (\ref{HiggsL}) that masses of the vector bosons must have a form of a product of the constant part of the Higgs field ($v$) and interaction constants.

The first relation (\ref{masses}) allows to connect the constant $v$ with  the Fermi constant $G_{F}$. In fact, the Fermi constant, which can be determined from the measurement of time of life of muon and from other CC data, is given by the expression
\begin{equation}\label{Fermiconst}
    \frac{G_{F}}{\sqrt{2}}=\frac{g^{2}}{8m^{2}_{W}}.
\end{equation}
From (\ref{masses}) and (\ref{Fermiconst}) we obviously have
\begin{equation}\label{Fermiconst1}
v^{2}=\frac{1}{\sqrt{2}G_{F}}.
\end{equation}
Thus, we find
\begin{equation}\label{Fermiconst1}
v=(\sqrt{2}G_{F})^{-1/2}\simeq 246~\mathrm{GeV}.
\end{equation}
The interaction constant $g$ is connected with the electric charge $e$ and the parameter $\sin\theta_{W}$
by the unification condition (\ref{unif}). From (\ref{unif}), (\ref{masses}) and  (\ref{Fermiconst1}) for the mass of the $W$ boson we find the following expression
\begin{equation}\label{mwmass}
 m_{W}=(\frac{\pi\alpha}{\sqrt{2}G_{F}})^{1/2}\frac{1}{\sin\theta_{W}},
\end{equation}
where $\alpha\simeq \frac{1}{137.036}$ is the fine-structure constant. For the mass of the $Z^{0}$ boson we have
\begin{equation}\label{mzmass}
m_{Z}=\frac{m_{W}}{\cos\theta_{W}}=(\frac{\pi\alpha}{\sqrt{2}G_{F}})^{1/2}\frac{1}{\sin\theta_{W}\cos\theta_{W}}.
\end{equation}
The parameter $\sin^{2}\theta_{W}$ can be determined from the data on the investigation of NC weak processes. From existing data  it was found the value $\sin^{2}\theta_{W}=0.23116(12)$  \cite{Beringer:1900zz}.

Thus, the Standard Model allows to connect masses of $W^{\pm}$ and $Z^{0}$ bosons with constants $G_{F}$, $\alpha$ and
$\sin^{2}\theta_{W}$.

For average of measured values of $m_{W}$ and $m_{Z}$ we have \cite{Beringer:1900zz}:
\begin{equation}\label{mW,Z}
 m_{W}=80.420 \pm 0.031 ~\mathrm{GeV},\quad  m_{Z}=91,1876\pm 0.0021  ~\mathrm{GeV}.
\end{equation}
Using the values of $G_{F}$, $\alpha$ and
$\sin^{2}\theta_{W}$ (and taking into account radiative corrections) for predicted by the SM values of $m_{W}$
and $m_{Z}$ we have
\begin{equation}\label{1mW,Z}
 m_{W}=80.381 \pm 0.014 ~\mathrm{GeV},\quad  m_{Z}=91,1874\pm 0.0021  ~\mathrm{GeV}.
\end{equation}
The agreement of the experimental data with one of the basic prediction of the SM is an important confirmation of the idea of the spontaneous breaking of the electroweak symmetry.

We will consider now the Higgs mechanism of the generation of masses of leptons and quarks. The fermion mass terms can be generated by a $SU_{L}(2)\times U_{Y}(1)$ invariant Yukawa Lagrangians. We will   consider first the charged leptons. The most general Yukawa Lagrangian which can generate the mass term of the charged leptons has the following  form
\begin{equation}\label{Lcharlep1}
\mathcal{L}^{lep}_{Y}=-\sqrt{2}\sum_{l_{1},l_{2}} \bar\psi^{lep}_{ l_{1} L}Y_{l_{1}l_{2}}l'_{2R}~\phi+\rm{h.c},
\end{equation}
where $Y$ is a $3\times3$ complex nondiagonal matrix. The Standard Model does not predict  elements of the matrix $Y$:
they are parameters of the SM.

After the spontaneous breaking of the symmetry from (\ref{lepSU2dub}),
(\ref{vev6}) and (\ref{Lcharlep1}) we have
\begin{equation}\label{Lcharlep2}
 \mathcal{L}^{lep}_{Y}=-\sum_{l_{1},l_{2}}\bar l'_{1L}Y_{l_{1}l'_{2}}l'_{2R}(v+H)+\rm{h.c}.
\end{equation}
The term proportional to $v$ is the  mass term of charged leptons. In order to present it in the canonical form we need to diagonalize  matrix   $Y$. The general complex matrix $Y$ can be diagonalized by the biunitary transformation
\begin{equation}\label{Lcharlep3}
    Y=V_{L}~y~V^{\dag}_{R},
\end{equation}
where $V_{L}$ and $V_{R}$ are unitary matrices and $y$ is a diagonal matrix with positive diagonal elements.
From (\ref{Lcharlep2}) and (\ref{Lcharlep3}) we find
\begin{equation}\label{Lcharlep4}
\mathcal{L}^{lep}_{Y}=-\sum_{l=e,\mu,\tau}\bar l_{L}m_{l}~l_{R}~(1+\frac{1}{v}H)+\rm{h.c}
=-\sum_{l=e,\mu,\tau} m_{l}~\bar l~l~(1+\frac{1}{v}H).
\end{equation}
Here
\begin{equation}\label{Lcharlep5}
 l_{L}=\sum_{l_{1}}( V^{\dag}_{L})_{l l_{1}}~l'_{1L},\quad l_{R}=\sum_{l_{1}}( V^{\dag}_{R})_{ll_{1}}~l'_{1R},\quad
l=l_{L}+l_{R}
\end{equation}
and
\begin{equation}\label{Lcharlep6}
    m_{l}=y_{l}~v.
\end{equation}
From (\ref{Lcharlep5}) it follows that $l(x)$ is the field of the charged leptons $l$  with  mass $m_{l}$ ($l=e,\mu,\tau$). Left-handed and right-handed components of the fields of leptons  with definite masses are connected, with primed left-handed fields, components of the doublets $\psi^{lep}_{l L}(x)$, and primed singlets right-handed fields $l'_{R}$  by the unitary transformations (\ref{Lcharlep5}).

The second term of  (\ref{Lcharlep4}) is the Lagrangian of  interaction of leptons and the Higgs boson
\begin{equation}\label{1Lcharlep}
\mathcal{L}_{Y}=-\sum_{l=e,\mu,\tau}f_{l}~\bar l~l~H,
\end{equation}
where dimensionless interaction constants $f_{l}$ are given by the relation
\begin{equation}\label{2Lcharlep}
f_{l}=\frac{1}{v}~m_{l}=(\sqrt{2}G_{F})^{1/2}m_{l}\simeq 4.06\cdot 10^{-3}
    \frac{m_{l}}{\mathrm{GeV}}.
\end{equation}
Let us express leptonic electromagnetic, charged and neutral currents in terms of the fields of leptons with definite masses $l(x)$. Taking into account the unitarity of the matrices $V_{L}$ and $V_{R}$ for the EM current we have
\begin{eqnarray}\label{1EM}
j^{EM}_{\alpha}&=&\sum_{l} (-1)\bar l'_{L}\gamma_{\alpha} l'_{L}+ \sum_{l} (-1)\bar l'_{R}\gamma_{\alpha} l'_{R}\nonumber\\&=&
\sum_{l} (-1)\bar l_{L}\gamma_{\alpha} l_{L}+ \sum_{l} (-1)\bar l_{R}\gamma_{\alpha} l_{R}=
\sum_{l} (-1)\bar l~\gamma_{\alpha}~ l.
\end{eqnarray}
For the leptonic charged current we find
\begin{equation}\label{1CC}
j^{CC}_{\alpha}=2\sum_{l}\bar \nu'_{lL}\gamma_{\alpha} l_{L}=2\sum_{l}\bar \nu_{lL}\gamma_{\alpha} l_{L},
\end{equation}
where
\begin{equation}\label{2CC1}
\nu_{lL}=\sum_{l_{1}}( V^{\dag}_{L})_{ll_{1}}~\nu'_{l_{1}L}.
\end{equation}
The field $\nu_{l}$ is called flavor neutrino field.

Finally, for the leptonic NC we obtain the following expression
\begin{eqnarray}
j^{NC}_{\alpha}&=&\sum_{l} \bar \nu'_{lL}\gamma_{\alpha}  \nu'_{lL}-\sum_{l}\bar l'_{L}\gamma_{\alpha} l'_{L}-2\sin^{2}\theta_{W}j^{EM}_{\alpha}  \\
   &=& \sum_{l} \bar \nu_{lL}\gamma_{\alpha}  \nu_{lL}-\sum_{l}\bar l_{L}\gamma_{\alpha} l_{L}-2\sin^{2}\theta_{W}j^{EM}_{\alpha}.
\end{eqnarray}
We will consider now briefly  the Brout-Englert-Higgs mechanism of the generation of masses of quarks. Let us assume that in the total Lagrangian enter the following $SU_{L}(2)\times U_{Y}(1)$ invariant Lagrangian of the Yukawa interaction of quark and Higgs fields
\begin{equation}\label{quarkH}
\mathcal{L}_{Y}^{\rm{quark}}=-\sqrt{2}\,\sum_{k,q_{1}=d,s,b}
\bar\psi_{kL}~Y_{k q_{1}}^{down}~ q'_{1R}~\phi -\sqrt{2}~\sum_{k,q_{1}=u,c,t
}\bar\psi _{kL}\, ~Y^{up}_{k q_{1}}\,q'_{1R} \,~ \tilde \phi+\rm{h.c.}
\end{equation}
Here
\begin{eqnarray}
\psi_{1L}=\left(
\begin{array}{c}
u'_{L} \\
d'_L\\
\end{array}
\right),~ \psi_{2 L}=\left(
\begin{array}{c}
c'_{L}\\
s'_L \\
\end{array}
\right),~ \psi_{3 L}=\left(
\begin{array}{c}
t'_L\\
b'_L\\
\end{array}
\right)\label{quarkSU2dub}
\end{eqnarray}
are quark doublets,
\begin{equation}\label{quarkH1}
\tilde{\phi}=i\,\tau_{2}\phi^{*}
\end{equation}
is the conjugated Higgs doublet and $Y_{k q_{1}}^{down}$,  $Y_{k q_{1}}^{up}$ are $3\times 3$ complex nondiagonal matrices.

After the spontaneous breaking of the symmetry in the unitary gauge we have
\begin{equation}\label{quarkH2}
\phi(x) ={  0   \choose \frac{v+H(x)}{\sqrt{2}}} ,\quad
\tilde \phi(x) ={  \frac{v+H(x)}{\sqrt{2}}   \choose 0}.
\end{equation}
From (\ref{quarkH}) and (\ref{quarkH2}) we find
\begin{eqnarray}\label{quarkH4}
\mathcal{L}_{Y}^{\rm{quark}}&=&-\sum_{q_{1},q_{2}=d,s,b}
\bar q'_{1L}~Y_{q_{1} q_{2}}^{down}~ q'_{2R}~(v+H)\nonumber\\
&-&\sum_{q_{1},q_{2}=u,c,t}
\bar q'_{1L}~Y_{q_{1} q_{2}}^{up}~ q'_{2R} \,~ (v+H)+\rm{h.c.}
\end{eqnarray}
For the complex matrices $Y^{down}$ and $Y^{up}$ we have
\begin{equation}\label{quarkH5}
Y^{down}=V^{down}_{L}~y^{down}~V^{down\dag}_{R},\quad Y^{up}=V^{up}_{L}~y^{up}~V^{up\dag}_{R}.
\end{equation}
Here $V^{down}_{L,R}$ and $V^{up}_{L,R}$ are unitary matrices and $y^{down}$, $y^{up}$ are diagonal matrices with positive diagonal elements.

Using (\ref{quarkH5}) for the Lagrangian $\mathcal{L}_{Y}^{\rm{quark}}$ we find
\begin{equation}\label{quarkH6}
\mathcal{L}_{Y}^{\rm{quark}}=-\sum_{q=u,d,c,s,t,b} m_{q}~\bar q~q~(1+\frac{1}{v}H).
\end{equation}
Here
\begin{equation}\label{quarkH7}
    m_{q}=y_{q}~v,\quad q=u,d,c,s,t,b
\end{equation}
are masses of the quarks,
\begin{equation}\label{quarkH8}
    q_{L}=\sum_{q_{1}=d,s,b}(V_{L}^{down\dag})_{q q_{1}}~q'_{1L}~~(q=d,s,b)\quad  q_{L}=\sum_{q_{1}=u,c,t}(V_{L}^{up\dag})_{q q_{1}}~q'_{1L}~~(q=u,c,t)
\end{equation}
and
\begin{equation}\label{quarkH9}
    q_{R}=\sum_{q_{1}=d,s,b}(V_{R}^{down\dag})_{q q_{1}}~q'_{1R}~~(q=d,s,b)\quad  q_{R}=\sum_{q_{1}=u,c,t}(V_{R}^{up\dag})_{q q_{1}}~q'_{1R}~~(q=u,c,t)
\end{equation}
The first terms in the r.h.s. of Eq. (\ref{quarkH6}) is the mass term of the quark
\begin{equation}\label{quarkH10}
\mathcal{L}_{m}^{\rm{quark}}=-\sum_{q=u,d,...} m_{q}\bar q~q.
\end{equation}
The second term
\begin{equation}\label{quarkH11}
\mathcal{L}_{H}^{\rm{quark}}= -\sum_{q=u,d
,...} f_{q}~\bar q q ~H
\end{equation}
is the Lagrangian of the interaction of quarks and the scalar Higgs boson. The interaction constants $f_{q}$
 are given by the relation
\begin{equation}\label{quarkH12}
    f_{q}=\frac{m_{q}}{v}=m_{q}(\sqrt{2}G_{F})^{1/2}\simeq 4.06\cdot 10^{-3}
    \frac{m_{q}}{\mathrm{GeV}}.
\end{equation}
Let us express  the electromagnetic current, neutral current and charged current of quarks in terms of the fields of quarks with definite masses. Taking into account the unitarity of the matrices $V^{up}_{L,R}$ and $V^{down}_{L,R}$ for the electromagnetic current of quarks we have the following expression
\begin{equation}\label{qcurrents1}
j^{\mathrm{EM}} _{\alpha}= \frac{2}{3}\sum_{q=u,c,t}\bar q'\gamma_{\alpha} q'+(-\frac{1}{3})\sum_{q=d,s,b}\bar q'\gamma_{\alpha} q'=\sum_{q=u,d,...}e_{q}~\bar q\gamma_{\alpha} q,
\end{equation}
where $e_{q}=\frac{2}{3}~\mathrm{for}~ q=u,c,t$ and $e_{q}=-\frac{1}{3}~\mathrm{for}~ q=d,s,b$.

Analogously, for the the neutral current of quarks we find
\begin{eqnarray}\label{qcurrents2}
j^{\rm{NC}}_{\alpha}&=&\sum_{q=u,c,t}\bar q'_{L}\gamma_{\alpha} q'_{L}-\sum_{q=d,s,b}\bar q'_{L}\gamma_{\alpha} q'_{L}-2\sin^{2}\theta_{W}j^{\mathrm{EM}} _{\alpha}\nonumber\\
&=&\sum_{q=u,c,t}\bar q_{L}\gamma_{\alpha} q_{L}-\sum_{q=d,s,b}\bar q_{L}\gamma_{\alpha} q_{L}-2\sin^{2}\theta_{W}j^{\mathrm{EM}} _{\alpha}.
\end{eqnarray}
Thus, NC of the Standard Model is diagonal over quark fields (conserves quark flavor).

Finally, for the charged current of quarks we have
\begin{equation}\label{qcurrents3}
 j^{\rm{CC}}_{\alpha}=\bar u'_{L}\gamma_{\alpha} d'_{L}+\bar c'_{L}\gamma_{\alpha} s'_{L}+\bar t'_{L}\gamma_{\alpha} b'_{L}=\bar u_{L}\gamma_{\alpha} d^{\mathrm{mix}}_{L}+\bar c_{L}\gamma_{\alpha} s^{\mathrm{mix}}_{L}+
 \bar t_{L}\gamma_{\alpha} b^{\mathrm{mix}}_{L}.
\end{equation}
Here
\begin{equation}\label{qcurrents4}
d^{\mathrm{mix}}_{L}=\sum_{q=d,s,b}V_{uq}q_{L},~~s^{\mathrm{mix}}_{L}=\sum_{q=d,s,b}V_{cq}q_{L},~~b^{\mathrm{mix}}_{L}=
\sum_{q=d,s,b}V_{tq}q_{L}.
\end{equation}
The matrix $V=V^{up}_{L}~V^{down\dag}_{L}$ is a unitary $3\times3$  Cabibbo-Kobayashi-Maskawa (CKM) matrix. Thus, the fields of down quarks enter into CC in the mixed form. The mixing is connected with the fact that the unitary matrices $V^{up}_{L}$ and $V^{down}_{L}$ are different.

The CKM matrix is characterized by three mixing angles $\theta_{12}$,
$\theta_{23}$, $\theta_{13}$ and one phase $\delta$ responsible for the $CP$ violation in the quark sector. It can be presented  in the same form as the neutrino mixing matrix (see (\ref{unitmixU1})). Existing data allows to determine all matrix elements of CKM matrix. From the global fit of the data of numerous experiments  it was found \cite{Beringer:1900zz}
\begin{eqnarray}
|V|=\left(\begin{array}{ccc}0.97427\pm 0.00015 &0.22534\pm 0.00065&0.00351\pm 0.00015\\
0.22520\pm 0.00065&
0.97344\pm 0.00016&0.0412^{+0.0011}_{-0.0005}\\
0.00867^{+0.00029}_{-0.00031}&
0.0404^{+0.0011}_{-0.0005}&0.999146^{+0.000021}_{-0.000046}
\end{array}\right)
\label{unitmixU2}
\end{eqnarray}
From (\ref{2Lcharlep}) and (\ref{quarkH12}) for the masses of charged leptons and quarks we have
\begin{equation}\label{lqmasses}
m_{l}=f_{l}v,\quad m_{q}=f_{q}v.
\end{equation}
Thus, masses of leptons (quarks) have the form of the product of constant $v$ (coming from the Higgs field) and the constants of interaction of leptons (quarks) and the Higgs bosons. Notice that masses of $W^{\pm}$  and $Z^{0}$ vector bosons have the same form (see (\ref{masses})).

Masses of leptons and quarks are known. From (\ref{lqmasses}) follows that {\em  the SM predicts the constants of interaction of leptons and quarks with the Higgs boson.} The first LHC measurements of the constants  $f_{\tau}$ and  $f_{b}$ are in agreement with the SM prediction (see \cite{Chatrchyan:2013zna,Chatrchyan:2014nva}).

Up to now we considered the Standard Model Brout-Englert-Higgs mechanism of the generation of masses of charged leptons and quarks. What about neutrinos? As we discussed earlier, in the minimal Standard Model there are no right-handed neutrino fields. Thus, in the minimal SM there is no Yukawa interaction  which can generate neutrino mass term. This means that {\em after spontaneous breaking of the electroweak symmetry neutrino fields in the SM remain two-component Weyl fields and  neutrino mass term can be generated only by a beyond the Standard Model mechanism.}

In conclusion we will present some additional arguments in favor of a beyond the SM origin of the neutrino masses. Let us assume that not only   $\nu'_{lL}$ but also   $\nu'_{lR}$
are Standard Model fields. In this case we  have the following $SU_{L}(2)\times U_{Y}(1)$ invariant  Yukawa interaction of lepton and Higgs fields
\begin{equation}\label{nuYukawa}
\mathcal{L}^{\nu}_{Y}=-\sqrt{2}\sum_{l'l}\overline  \psi^{lep}_{l_{1}L}Y^{\nu}_{l_{1}l_{2}}\nu'_{l_{2}R}\tilde{\phi }+\mathrm{h.c.}.
\end{equation}
After spontaneous breaking of the electroweak symmetry from (\ref{nuYukawa}) we obtain the Dirac neutrino mass term
\begin{equation}\label{Dmass}
 \mathcal{L}^{\mathrm{D}}=-v\sum_{l',l}\bar\nu'_{l'L}\,Y^{\nu}_{l'l} \nu'_{lR}
+\mathrm{h.c.}=-\sum_{l',l}\bar\nu_{l'L}~M_{l'l}^{D}\nu_{lR}+\mathrm{h.c.}
\end{equation}
Here $M^{D}=vV^{\dag}_{L}Y^{\nu}$ where the matrix $V_{L}$ connects fields $\nu'_{lL}$ and flavor neutrino fields  $\nu_{lL}$ (see (\ref{2CC1})).
After the standard diagonalization of the matrix $V^{\dag}_{L}Y^{\nu}$  we find
 \begin{equation}\label{Dmass1}
\mathcal{L}^{\mathrm{D}}=\sum^{3}_{i=1}m_{i}\bar \nu_{i}\nu_{i} ,\quad \nu_{lL}=\sum_{i} U_{li}\nu_{iL},
 \end{equation}
where $U$ is a unitary mixing matrix and $\nu_{i}$ is a field of the Dirac neutrinos with mass $m_{i}$. For
neutrino mass we have
\begin{equation}\label{Dmass2}
    m_{i}=v~y_{i},
\end{equation}
where $y_{i}$ is the Yukawa  coupling constant.

In order to estimate  $y_{i}$ we need to know neutrino masses. Values of neutrino masses are determined by the lightest neutrino mass $m_{0}$ which is unknown at present. We will consider two extreme cases
\begin{enumerate}
  \item Normal mass hierarchy $m_{1}\ll m_{2}\ll m_{3}$, $m_{1}\ll \sqrt{\Delta m^{2}_{S}}\simeq 9\cdot 10^{-3}~\mathrm{eV}$,
 \begin{center}
      $y_{1}\ll y_{2}\ll y_{3}\simeq \frac{\sqrt{\Delta m^{2}_{A}}}{v}\simeq 2\cdot 10^{-13}$.
\end{center}
\item Inverted mass hierarchy $m_{3}\ll m_{1}< m_{2}$, $m_{3}\ll \sqrt{\Delta m^{2}_{A}}\simeq 5\cdot 10^{-2}~\mathrm{eV}$,
\begin{center}
    $ y_{3}\ll y_{1}\lesssim y_{3}\simeq \frac{\sqrt{\Delta m^{2}_{A}}}{v}\simeq 2\cdot 10^{-13}$.
\end{center}
\item Quasi-degenerate mass spectrum $m_{1,3}\gg \sqrt{\Delta m^{2}_{A}}\simeq 5\cdot 10^{-2}~\mathrm{eV}$.

In this case $m_{1}\lesssim m_{2}\lesssim m_{3}\simeq \frac{\sum m_{i}}{3}~~~ \mathrm{or}~~~ m_{1}\lesssim m_{2}\lesssim m_{3}\simeq m_{\beta}$.

Using (\ref{MainzTroitsk}) and (\ref{sumnu}) from the cosmological data and tritium $\beta$-decay data we find, respectively,
\begin{center}
$y_{1}\lesssim y_{2}\lesssim y_{3}\simeq 3\cdot 10^{-13}, \quad y_{1}\lesssim y_{2}\lesssim y_{3}\simeq  10^{-11}$
\end{center}
\end{enumerate}
Values of the quarks and leptons Yukawa coupling constants  depend on generation. For the particles of the first, second and third generation they  are of the order $ 10^{-6}-10^{-5}$, $ 10^{-3}-10^{-2}$ and $ 10^{-2}-1$, respectively. Thus, neutrino Yukawa coupling constants are many orders of magnitude smaller than
Yukawa constants of quarks and leptons. Extremely small neutrino masses and, correspondingly, neutrino Yukawa coupling constants  are   an evidence that {\em masses of quarks, leptons and neutrinos are not of the same SM origin.}

\section{Beyond the Standard Model neutrino masses}

In the Standard Model with left-handed, two-component Weyl fields $\nu_{lL}$ the neutrino mass term can not be generated. The neutrino mass term can be generated only by a beyond the SM mechanism. There are many approaches to neutrino masses
(see \cite{Mohapatra:2006gs}). {\em The most economical possibility of generation of neutrino masses and mixing is provided by an effective Lagrangian.}

The  effective Lagrangian method \cite{Weinberg:1979sa,Wilczek:1979hc} is a general, powerful method which allows to describe effects of a beyond the SM physics in the electroweak region. The effective Lagrangian is a  $SU_{L}(2)\times U_{Y}(1)$ invariant, nonrenormalizable  Lagrangian
 built from SM fields ({\em including Higgs field}). It has the following form
\begin{equation}\label{effLag}
 \mathcal{L}_{4+n}^{\mathrm{eff}}=\sum_{n=1,2,...}  \frac{O_{4+n}}{\Lambda^{n}} +\mathrm{h.c.}.
\end{equation}
Here $ O_{4+n}$ is a $SU_{L}(2)\times U_{Y}(1)$ invariant operator which has dimension $M^{4+n}$ and $\Lambda$ is a constant of the dimension $M$. The constant  $\Lambda$ characterizes a scale of a new, beyond the SM physics.

In order to generate the neutrino mass term we need to build the effective Lagrangian which is quadratic in the lepton fields. The terms
$\bar \psi^{lep}_{lL}\tilde{\phi }$  and $\tilde{\phi }^{\dag}\psi^{lep}_{lL}$                              ($l=e,\mu,\tau$) are $SU_{L}(2)\times U_{Y}(1)$ invariants which have dimensions $M^{5/2}$. After spontaneous breaking of the symmetry they contain, correspondingly, $v\bar\nu'_{lL}$ and $v\nu'_{lL}$. {\em The effective Lagrangian which generate neutrino mass term has the following lepton-number violating form} \cite{Weinberg:1979sa}
\begin{equation}\label{effective}
 \mathcal{L}_{5}^{\mathrm{eff}}=-\frac{1}{\Lambda}~\sum_{l_{1},l_{2}}(\bar \psi^{lep}_{l_{1}L}\tilde{\phi })~ Y'_{l_{1}l_{2}}~(\tilde{\phi }^{T} (\psi^{lep}_{l_{2}L})^{c})+\mathrm{h.c.}.
\end{equation}
Here $Y'=(Y')^{T}$ is a symmetric dimesionless $3\times3$ matrix and $\Lambda$ is a parameter which
characterizes a scale of a  beyond the SM  lepton number violating physics.

After spontaneous breaking of the electroweak symmetry  from (\ref{quarkH2}) and (\ref{effective}) we find
\begin{equation}\label{effective1}
\mathcal{L}_{I}^{\mathrm{eff}}=-\frac{1}{2\Lambda} \sum_{l_{1},l_{2}}
 \bar\nu'_{l_{1}L}~ Y'_{l_{1}l_{2}}~ (\nu'_{l_{2}L})^{c}~(v+H)^{2}+\mathrm{h.c.}
\end{equation}
The term proportional to $v^{2}$ is the neutrino mass term.

The flavor neutrino fields $\nu_{lL}$, which enter into the leptonic charged and neutral currents, are connected with fields $\nu'_{lL}$ by the relation (\ref{2CC1}). In terms of the flavor neutrino fields from (\ref{effective1}) we obtain the Majorana mass term (\ref{1Mjmass}) in which the matrix  $M^{M}$ is given by the following expression
\begin{equation}\label{Mjmat}
M^{M}=\frac{v^{2}}{\Lambda}~ Y,
\end{equation}
where
\begin{equation}\label{Mjmat1}
Y=V^{\dag}_{L}Y'(V^{\dag}_{L})^{T}
\end{equation}
is a symmetrical $3\times 3$ matrix. We have
\begin{equation}\label{1Mjmat1}
  Y=U~ y ~U^{T},
\end{equation}
where $U^{\dag}U=1$ and $ y_{ik}= y_{i}~\delta_{ik},~~y_{i}>0$.

From (\ref{Mjmat}) and (\ref{1Mjmat1}) for the Majorana neutrino mass we find the following expression
\begin{equation}\label{Mjmat2}
m_{i}=\frac{v}{\Lambda}~( y_{i}v),\quad i=1,2,3.
\end{equation}
Majorana neutrino mass $m_{i}$, generated by the effective Lagrangian (\ref{effective}), is
a product of a "typical fermion mass" $v~ y_{i}$ and a suppression factor which is given by the ratio of the electroweak scale $v$ and a scale $\Lambda$ of a  lepton-number violating physics ($\Lambda\gg v$). Thus, effective Lagrangian approach provides a natural framework for the generation of neutrino masses which are much smaller than the masses of leptons and quarks. Let us stress that such a scheme does not put any constraints of the mixing matrix $U$.

In order to estimate the parameter $\Lambda$ we need to know neutrino masses $m_{i}$ and Yukawa coupling constant $ y_{i}$. Let us  assume hierarchy of neutrino masses $m_{1}\ll m_{2}\ll m_{3}$. For the mass of the heaviest neutrino  we have in this case $m_{3}\simeq \sqrt{\Delta m^{2}_{A}}\simeq 5\cdot 10^{-2}$ eV. Assuming also that $ y_{3}$ is of the order of one, we find the following estimate $\Lambda\simeq 10^{15}$ GeV. Thus, small Majorana neutrino masses could be a signature of a  very large lepton number violating scale in physics.\footnote{Let us stress that for the dimensional arguments we used  it is important that Higgs is not composite particle and {\em exist scalar Higgs field having dimension $M$}. Discovery of the Higgs boson at CERN \cite{Aad:2012tfa,Chatrchyan:2012ufa} confirm this assumption.}

Effective Lagrangian  (\ref{effective}) could be a result of exchange of virtual super-heavy  Majorana leptons between lepton-Higgs pairs \cite{Weinberg:1980bf}.\footnote{An example of the effective Lagrangian is the  Fermi Lagrangian which describe $\beta$-decay and other low-energy processes. This effective Lagrangian is generated by the exchange of the virtual $W$-boson between $e-\nu$ and $p-n$ pairs. It is a product of the Fermi constant which has dimension $M^{-2}$ and dimension six four-fermion operator.}

In fact,  let us assume that exist heavy Majorana leptons $N_{i}$ ($i=1,2,...N$), singlets of $SU_{L}(2)\times U_{Y}(1)$ group, which have the following Yukawa lepton-number violating interaction
\begin{equation}\label{HeavyMj}
 \mathcal{L}_{I}^{Y}=-\sqrt{2}\sum _{l,i} \bar \psi^{lep}_{lL}\tilde{\phi }y'_{li} ~N_{iR}+\mathrm{h.c.}.
\end{equation}
Here  $y'_{li}$ are dimensionless Yukawa coupling constants and  $N_{i}=N^{c}_{i}$ is the Majorana field  with mass $M_{i}$ ($M_{i}\gg v$).

In the second order of the perturbation theory with virtual $N_{i}$   at the electroweak energies ($Q^{2}\ll M^{2}_{i}$) the interaction (\ref{HeavyMj}) generates the following effective Lagrangian
\begin{equation}\label{HeavyMj1}
 \mathcal{L}^{\mathrm{eff}}=-\sum_{l',l}(\bar \psi^{lep}_{l'L}\tilde{\phi })~ (\sum_{i}y'_{l'i}\frac{1}{M_{i}}y'_{li})~(\tilde{\phi }^{T} (\psi^{lep}_{l L})^{c})+\mathrm{h.c.}.
\end{equation}
After spontaneous breaking of the electroweak symmetry from (\ref{HeavyMj1}) we obtain Majorana neutrino mass term
\begin{equation}\label{HeavyMj2}
\mathcal{L}^{\mathrm{L}}=-\frac{1}{2}\sum_{l',l}\bar \nu'_{l'L}~ (\sum_{i}y'_{l'i}\frac{v^{2}}{M_{i}}y'_{l i})~ (\nu'_{l L})^{c}+\mathrm{h.c.}.
\end{equation}
In terms of  flavor neutrino fields $\nu_{l L}$  from (\ref{HeavyMj2}) we find
\begin{equation}\label{HeavyMj3}
\mathcal{L}^{\mathrm{L}}=-\frac{1}{2}\sum_{l',l}\bar \nu_{l'L}~ M_{l'l}^{L}~ (\nu_{l L})^{c}+\mathrm{h.c.}.
\end{equation}
Here
\begin{equation}\label{HeavyMj4}
M^{L}=y~\frac{v^{2}}{M}~y^{T}
\end{equation}
where $y=V_{L}^{\dag}y'$. From (\ref{1Mjmass1}) and (\ref{HeavyMj4}) for the Majorana neutrino mass $m_{i}$ ($i=1,2,3$) we find the following expression
\begin{equation}\label{HeavyMj5}
 m_{i}=\sum^{N}_{k=1} (U^{\dag}y)^{2}_{ik}~ \frac{v^{2}}{M_{k}}.
\end{equation}
The scale of a new lepton-number violating physics is determined by masses of heavy Majorana leptons $N_{i}$. It follows from (\ref{HeavyMj5}) that Majorana neutrino masses are suppressed with respect to the masses of other fundamental fermions by the factors $\frac{v}{M_{k}}\ll 1$.

Let us summarize our discussion of the generation of the neutrino masses by the Weinberg effective Lagrangian.

 \begin{enumerate}
   \item There is one possible lepton number violating effective Lagrangian.  After spontaneous breaking of the symmetry it leads  to the Majorana neutrino mass term which is the  only possible (in the case of the left-handed fields $\nu_{lL}$) neutrino mass term (see \cite{Gribov:1968kq}). Neutrino masses in this approach are suppressed with respect to the masses of lepton and quarks by the ratio of the electroweak scale $v$ and a scale $\Lambda$ of a new lepton-number violating physics ($\Lambda\gg v$).  The Lagrangian (\ref{effective}) is the only effective Lagrangian of the dimension 5 (proportional to $\frac{1}{\Lambda}$).  This means that {\em neutrino masses are the most sensitive probe of a new physics at a scale which is much larger than the electroweak scale.}

   \item  Number of Majorana neutrinos with definite masses is determined by the number of lepton flavors and is equal to three.

\item  Heavy Majorana leptons with masses much larger than $v$ could exist.

 \end{enumerate}

Alternative mechanism of the generation of small Majorana neutrino masses is the famous seesaw mechanism \cite{ Minkowski:1977sc,Yanagida:1979as,GellMann:1980vs,Glashow:1979nm,Mohapatra:1980ia}. This mechanism is based of GUT models (like $SO(10)$) with multiplets  which contain not only left-handed neutrino fields $\nu_{lL}$ but also right-handed fields. In such models the most general lepton-number violating Dirac and Majorana mass term (\ref{1DMmass}) is generated. If we assume that
\begin{enumerate}
  \item $M^{L}=0$.
  \item The elements of the matrix $M^{D}$ are proportional to $v$ (the Dirac term  $M^{D}$ is generated by the standard Higgs mechanism)
  \item The right-handed Majorana term $M^{R}$ (which can be always diagonalized) is given by  $M_{ik}^{R}=M_{k}\delta_{ik},~~~M_{k}\gg v$
\end{enumerate}
then we come to the Majorana neutrino mass term
\begin{equation}\label{seesawMj}
\mathcal{L}^{\mathrm{L}}= -\frac{1}{2}\,\sum_{l',l}
\bar\nu_{l'L}\,(M_{l'l}^{L})_{\mathrm{seesaw}} ( \nu_{lL})^{c}
+\mathrm{h.c.},
\end{equation}
where
\begin{equation}\label{seesawMj1}
(M^{L})_{\mathrm{seesaw}} =-M ^{D}~ (M^{R})^{-1}~ (M ^{D})^{T}.
\end{equation}
  In the seesaw case in the mass spectrum there are three light (neutrino) Majorana masses $m_{i}$ and heavy lepton
Majorana masses $M_{k}$. From (\ref{seesawMj1}) it follows that the scale of neutrino masses is determined by the factor $\frac{v^{2}}{M_{k}}\ll v$.

 The seesaw mechanism of the generation of the neutrino masses is equivalent to the effective Lagrangian mechanism considered before. Let us notice that the mechanism based on the interaction (\ref{HeavyMj}) is called type I seesaw. The effective Lagrangian (\ref{effective}) can also be generated by the Lagrangian of interaction of lepton-Higgs doublets with a heavy triplet leptons (type III seesaw) and by the Lagrangian of interaction of lepton doublets and Higgs doublets with heavy triplet scalar bosons (type II seesaw).

\section{Implications of the standard seesaw mechanisms of neutrino mass generation}

In this section we will briefly discuss practical implications  of  the effective Lagrangian (seesaw) mechanism of the neutrino mass generation.

\begin{center}
\bf{Neutrinoless double $\beta$-decay }
\end{center}
The search for neutrinoless double $\beta$-decay ($0\nu\beta\beta$-decay)
\begin{equation}\label{betabeta}
(A,Z)\to (A,Z+2)+ e^{-}+e^{-}.
\end{equation}
of $^{76}\mathrm{Ge}$, $^{130}\mathrm{Te}$, $^{136}\mathrm{Xe}$ and other even-even nuclei is the most practical way which allows to reveal the nature of neutrinos with definite masses (Majorana or Dirac?)(see \cite{Doi:1985dx,Bilenky:1987ty,Avignone:2007fu,Bilenky:2014uka}).

The expected half-life of this process is extremely large (many orders of magnitude larger than time of life of the Universe). There are two main reasons for that.
\begin{enumerate}
  \item The process (\ref{betabeta}) is the second order of the perturbation theory process with the exchange of the virtual neutrinos between $n\to pe^{-}$ vertexes. The matrix element of the process is proportional to $G^{2}_{F}$.
  \item Because in the Hamiltonian of the standard weak interaction enter left-handed  neutrino fields
\begin{equation}\label{betabeta1}
    \nu_{eL}=\sum_{i}U_{ei}\nu_{iL}
\end{equation}
neutrino propagator has the form
\begin{equation}\label{betabeta2}
    \sum_{i}U^{2}_{ei}\frac{1-\gamma_{5}}{2}~\frac{\gamma\cdot q+m_{i}}{q^{2}-m^{2}_{i}}~\frac{1-\gamma_{5}}{2}\simeq \frac{m_{\beta\beta}}{q^{2}}~\frac{1-\gamma_{5}}{2}.
\end{equation}
Here
\begin{equation}\label{betabeta3}
m_{\beta\beta}=\sum_{i}U^{2}_{ei}m_{i}
\end{equation}
is the effective Majorana mass and $q$ is momentum of virtual neutrinos. From neutrino data it follows that $|m_{\beta\beta}|\lesssim 1$ eV. An average momentum of the virtual neutrino is about 100 MeV \cite{Doi:1985dx,Bilenky:1987ty}. Thus, the factor
$\frac{m_{\beta\beta}}{q^{2}}$ gives strong  suppression of the matrix element of $0\nu\beta\beta$-decay\footnote{It follows from (\ref{betabeta2}) that for massless neutrinos
$0\nu\beta\beta$-decay is forbidden. This corresponds to the theorem on the equivalence of theories with massless
Dirac and Majorana neutrinos \cite{Ryan-Okubo-NCS-2-234-1964,Case:1957zza}}
\end{enumerate}
In the case of  Majorana neutrino mixing (\ref{betabeta1}) the half-life of the $0\nu\beta\beta$-decay $T^{0\nu}_{1/2}(A,Z)$ have the following general form (see \cite{Doi:1985dx,Bilenky:1987ty})
\begin{equation}\label{betabeta4}
\frac{1}{T^{0\nu}_{1/2}(A,Z)}=|m_{\beta\beta}|^{2}~|M^{0\nu}(A,Z)|^{2}~
G^{0\nu}(Q,Z).
\end{equation}
Here $M^{0\nu}(A,Z)$ is the nuclear matrix element (NME), which is determined by the nuclear properties and does not depend on elements of the neutrino mixing matrix and small neutrino masses, and $G^{0\nu}(Q,Z)$ is known phase space factor which includes the Fermi function describing final state Coulomb interaction of two electrons and nuclei.

The calculation of NME is a very complicated many-body nuclear problem. At present NME for the $0\nu\beta\beta$-decay  of $^{76}\mathrm{Ge}$, $^{130}\mathrm{Te}$, $^{136}\mathrm{Xe}$  and other nuclei were calculated in the framework of
 NSM, QRPA, IBM, EDF and PHFB many-body approximate schemes (see review \cite{Bilenky:2014uka} and references thereby).
Results of these calculations  are significantly different. In the Table~\ref{tab:nme}  we present ranges of NME for $^{76}\mathrm{Ge}$ and other nuclei, ratios of maximal and minimal values of NME and ranges of  half-lives calculated under the assumption that $|m_{\beta\beta}|=0.1$ eV (see \cite{Bilenky:2014uka} for details).
\begin{table}[b!]
\caption{\label{tab:nme}
Ranges of calculated values of $|M^{0\nu}|$,
ratios
$
|M^{0\nu}|_{\text{max}} / |M^{0\nu}|_{\text{min}}
$
and
ranges of half-lives (calculated for $m_{\beta\beta} = 0.1 \, \text{eV}$)
for  the neutrinoless double $\beta$-decay of several nuclei of experimental interest.}
\begin{center}
\begin{tabular}{cccc}
$0\nu\beta\beta$-decay	&$|M^{0\nu}|$ &$\dfrac{|M^{0\nu}|_{\text{max}}}{|M^{0\nu}|_{\text{min}}}$ &$\displaystyle{T^{0\nu}_{1/2}(m_{\beta\beta} = 0.1 \, \text{eV}) \atop [10^{26}\,\text{y}]}$ \\
\hline
$ ^{76}\mathrm{Ge} \to ^{76}\mathrm{Se} $	&$3.59-10.39$ &$2.9$ &$1.0-8.6$ \\
$ ^{100}\mathrm{Mo}\to ^{100}\mathrm{Ru}$	&$4.39-12.13$ &$2.8$ &$0.1-0.8$ \\
$ ^{130}\mathrm{Te}\to ^{130}\mathrm{Xe}$	&$2.06-8.00$ &$3.9$ &$0.3-4.3$ \\
$ ^{136}\mathrm{Xe}\to ^{136}\mathrm{Ba}$	&$1.85-6.38$ &$3.4$ &$0.4-5.2$ \\
\hline
\end{tabular}
\end{center}
\end{table}

Up to now $0\nu\beta\beta$-decay was not observed
and rather stringent lower bounds on half-life of the $0\nu\beta\beta$-decay of different nuclei were obtained.
We will present here some recent results.

In the  EXO-200 experiment \cite{Albert:2014awa} the $0\nu\beta\beta$-decay of $^{136}\mathrm{Xe}$ (with 80.6\% enrichment in $^{136}\mathrm{Xe}$)  was searched for in the liquid  time-projection chamber. After $100~ kg\cdot y$ exposure the following lower bound was obtained
\begin{equation}\label{Exo}
    T^{0\nu}_{1/2}(^{136}\mathrm{Xe})>1.1\cdot 10^{25}~\mathrm{y}\quad (90\% CL)
\end{equation}
Using different calculations of NME from this result for the effective Majorana mass the following upper bounds were found
\begin{equation}\label{Exo1}
|m_{\beta\beta}|<(1.9-4.5)\cdot 10^{-1}~\mathrm{eV}
\end{equation}
In the KamLAND-Zen experiment \cite{TheKamLAND-Zen:2014lma} 383 kg of liquid $^{136}\mathrm{Xe}$ (enriched to 90.77\% ) was loaded in the liquid scintillator. After 115 days of exposure for the half-life of $^{136}\mathrm{Xe}$ the following lower bound was inferred
\begin{equation}\label{Kzen}
    T^{0\nu}_{1/2}(^{136}\mathrm{Xe})>1.3\cdot 10^{25}~\mathrm{y}\quad (90\% CL)
\end{equation}
Combining this result with the result of the previous run,  for the half-life of $^{136}\mathrm{Xe}$ it was obtained
\begin{equation}\label{Kzen1}
    T^{0\nu}_{1/2}(^{136}\mathrm{Xe})>2.6\cdot 10^{25}~\mathrm{y}\quad (90\% CL)
\end{equation}
From this bound for the effective Majorana mass it was found
\begin{equation}\label{Exo1}
|m_{\beta\beta}|<(1.4-2.8)\cdot 10^{-1}~\mathrm{eV}.
\end{equation}
In the germanium GERDA experiment \cite{Agostini:2013mzu} the $0\nu\beta\beta$-decay of $^{76}\mathrm{Ge}$ was studied. In the Phase-I of the experiment the germanium target mass was 21.6 kg (86\% enriched in $^{76}\mathrm{Ge}$). Very law background ($ 10^{-2}~cts/\mathrm{KeV}~kg~y$) was reached. For the
the lower bound of the half-life of $^{76}\mathrm{Ge}$ it was obtained the value\footnote{This result allowed  to refute the claim of the observation of the
$0\nu\beta\beta$-decay of $^{76}\mathrm{Ge}$ made in \cite{KlapdorKleingrothaus:2004wj}}
\begin{equation}\label{Gerda}
    T^{0\nu}_{1/2}(^{76}\mathrm{Ge})>2.1\cdot 10^{25}~\mathrm{y}\quad (90\% CL).
\end{equation}
Combining (\ref{Gerda}) with the results of Heidelberg-Moscow \cite{Klapdor-Kleingrothaus:2001yx} and IGEX \cite{Aalseth:2002rf} experiments it was found
\begin{equation}\label{Gerda1}
    T^{0\nu}_{1/2}(^{76}\mathrm{Ge})>3.0\cdot 10^{25}~\mathrm{y}\quad (90\% CL).
\end{equation}
From this bound for the effective Majorana mass it was obtained the following bound
\begin{equation}\label{Gerda3}
|m_{\beta\beta}|<(2-4)\cdot 10^{-1}~\mathrm{eV}.
\end{equation}
The value of the effective Majorana mass strongly depend on the character of neutrino mass spectrum. Two mass spectra are of the special interest.
\begin{enumerate}
  \item Normal hierarchy of neutrino masses ($m_{1}\ll m_{2}\ll m_{3}$).

In this case
\begin{equation}\label{NH1}
m_{2}\simeq \sqrt{\Delta m^{2}_{S}},~~~m_{3}\simeq \sqrt{\Delta m^{2}_{A}},~~~m_{1} \ll \sqrt{\Delta m^{2}_{S}}\simeq 8.7 ~10^{-3}~\mathrm{eV}
\end{equation}
and for the effective Majorana mass we find
\begin{equation}\label{NH2}
|m_{\beta\beta}|=|\sin^{2}\theta_{12}~e^{2i\alpha} \sqrt{\Delta m^{2}_{S}}+\sin^{2}\theta_{13}~ \sqrt{\Delta m^{2}_{A}}|
\end{equation}
where $2\alpha$ is the relative phase. Using best-fit values of the parameters we find
\begin{equation}\label{NH3}
\sin^{2}\theta_{12}~ \sqrt{\Delta m^{2}_{S}} \simeq 3\cdot 10^{-3}~\mathrm{eV},~~~ \sin^{2}\theta_{13}~ \sqrt{\Delta m^{2}_{A}} \simeq 1\cdot 10^{-3}~\mathrm{eV}
\end{equation}
From (\ref{NH2}) and (\ref{NH3}) we find the following upper bound
\begin{equation}\label{NH4}
|m_{\beta\beta}|\lesssim  4\cdot 10^{-3}~\mathrm{eV}.
\end{equation}
This bound is too small to be reached in the next generation of experiments on the search for $0\nu\beta\beta$-decay.

\item Inverted hierarchy of neutrino masses ($m_{3}\ll m_{1} < m_{2}$).

In this case for the neutrino masses we have
\begin{equation}\label{IH1}
m_{1}\simeq m_{2} \simeq \sqrt{\Delta m^{2}_{A}},~~~m_{3} \ll \sqrt{\Delta m^{2}_{A}}\simeq 5 ~10^{-2}~\mathrm{eV}.
\end{equation}
and the effective Majorana mass is equal to
\begin{equation}\label{IH2}
|m_{\beta\beta}|=\sqrt{\Delta m^{2}_{A}}(1-\sin^{2}2\theta_{12}~\sin^{2}\alpha )^{1/2}.
\end{equation}
Thus, we have
\begin{equation}\label{IH3}
\sqrt{\Delta m^{2}_{A}}\cos2\theta_{12} \leq  |m_{\beta\beta}|\leq \sqrt{\Delta m^{2}_{A}}.
\end{equation}
From this inequality it follows that in the case of the inverted hierarchy of the neutrino masses the value of the effective Majorana mass lies in the range
\begin{equation}\label{IH4}
2\cdot 10^{-2} \lesssim  |m_{\beta\beta}|\lesssim 5\cdot 10^{-2}~\mathrm{eV}.
\end{equation}
More detailed calculations (see, for example, \cite{Bilenky:2014uka}) shows that this result is valid for the inverted mass spectrum at
$m_{3}\leq 1\cdot 10^{-2}~\mathrm{eV}$.
\end{enumerate}
The aim of the future experiments on the search for $0\nu\beta\beta$-decay is to probe predicted by the inverted hierarchy of neutrino masses range (\ref{IH4}).

\begin{center}
\bf{On  the search for transitions into sterile neutrinos }
\end{center}
All data of atmospheric, solar, reactor and accelerator neutrino oscillation experiments are perfectly described
by three-neutrino mixing with   two neutrino mass-squared differences $\Delta m_{S}^{2}\simeq 7.5 \cdot 10^{-5}~\mathrm{eV}^{2}$ and $\Delta m_{A}^{2}\simeq 2.4 \cdot 10^{-5}~\mathrm{eV}^{2}$. Exist, however, indications in favor of neutrino oscillations with much larger neutrino mass-squared difference(s) about 1 $\mathrm{eV}^{2}$. These indications were obtained in the following short baseline neutrino experiments.
\begin{enumerate}
  \item The LSND \cite{Aguilar:2001ty} and MiniBooNE \cite{Aguilar-Arevalo:2013pmq,Conrad:2013mka} experiments.  In the LSND experiment neutrinos are produced in decays at rest of $\pi^{+}$'s and $\mu^{+}$'s. Electron antineutrinos, presumably produced in the transition $\bar\nu_{\mu}\to \bar\nu_{e}$, were detected. In the MiniBooNE experiment low energy excess of
 $\nu_{e}$ ( $\bar\nu_{e}$) was observed in  the $\nu_{\mu}$ ($\bar\nu_{\mu}$) experiments.

\item Reactor neutrino experiments. Indications in favor of disappearance of the reactor $\bar\nu_{e}$'s were obtained from the  new analysis  of the data of old reactor neutrino experiments \cite{Mention:2011rk}
     in which recent calculations of
  the reactor neutrino flux  \cite{Mueller:2011nm,Huber:2011wv} was used.

  \item Radiative source experiments. In the calibration experiments, performed with radiative sources by the GALLEX  \cite{Kaether:2010ag} and SAGE \cite{Abdurashitov:2009tn} collaborations, a deficit of  $\nu_{e}$'s was observed.
\end{enumerate}
In order to interpret these  data  in terms of neutrino oscillations we must assume that exist more than three neutrinos with definite masses and in addition to the flavor $\nu_{e},\nu_{\mu},\nu_{\tau}$ exist also sterile neutrinos.

In the case of the simplest 3+1 scheme with three light neutrinos and one neutrino with mass about 1 eV  for short baseline experiments, sensitive to large $\Delta m^{2}_{14}$, from (\ref{Genexp5})  we find the following expression for $\nua{\alpha}\to \nua{\alpha'}$ transition probability
\begin{equation}\label{transition1}
P(\nua{\alpha}\to \nua{\alpha'})=\delta_{\alpha,\alpha'}-4(\delta_{\alpha,\alpha'}-|U_{\alpha' 4}|^{2})|U_{\alpha' 4}|^{2}\sin^{2}{\frac{\Delta m^{2}_{14}L}{4E}},
\end{equation}
where  $\Delta m^{2}_{14}=m^{2}_{4}-m^{2}_{1}$.

 From this expression for $\nua{\mu}\to \nua{e}$ appearance probability and $\nua{e}\to \nua{e}$ and $\nua{\mu}\to \nua{\mu}$ disappearance probabilities we have, respectively, the following expressions
\begin{equation}\label{transition2}
P(\nua{\mu}\to \nua{e})=\sin^{2}2\theta_{e\mu}\sin^{2}{\frac{\Delta m^{2}_{14}L}{4E}},
\end{equation}
\begin{equation}\label{transition3}
P(\nua{e}\to \nua{e})=1-\sin^{2}2\theta_{e e}\sin^{2}{\frac{\Delta m^{2}_{14}L}{4E}},
\end{equation}
and
\begin{equation}\label{transition4}
P(\nua{\mu}\to \nua{\mu})=1-\sin^{2}2\theta_{\mu\mu}\sin^{2}{\frac{\Delta m^{2}_{14}L}{4E}}.
\end{equation}
Here
\begin{equation}\label{transition5}
\sin^{2}2\theta_{e\mu}=4|U_{e 4}|^{2}|U_{\mu 4}|^{2},~\sin^{2}2\theta_{e e}=4|U_{e 4}|^{2}(1-|U_{e 4}|^{2},~
\sin^{2}2\theta_{\mu\mu}=4|U_{\mu 4}|^{2}(1-|U_{\mu 4}|^{2}.
\end{equation}
The Global  analysis of all existing short baseline neutrino data was performed recently in \cite{Giunti:2013aea,Kopp:2013vaa}. These analysis reveal inconsistency (tension) of existing short baseline data. The reason for this tension is connected with the fact that the amplitudes of the oscillations are constrained by the relation
\begin{equation}\label{transition6}
\sin^{2}2\theta_{e\mu}\simeq \frac{1}{4}~\sin^{2}2\theta_{e e}~ \sin^{2}2\theta_{\mu\mu},
 \end{equation}
which can be easily obtained from (\ref{transition5}), if we take into account that $ |U_{e 4}|^{2}\ll 1$ and
 $ |U_{\mu 4}|^{2}\ll 1$.

Allowed regions of the parameters $\sin^{2}2\theta_{e\mu}$ and  $\sin^{2}2\theta_{e e}$, determined by
 $\nua{\mu}\to \nua{e}$ and $\nua{e}\to \nua{e}$ data, requires disappearance of $\nua{\mu}$ (due to the  constraint (\ref{transition6})). However, there are no indications in favor of $\nua{\mu}\to \nua{\mu}$  disappearance in  short baseline experiments \cite{Dydak:1983zq,Adamson:2011ch,Cheng:2012yy}.

Notice that in more complicated neutrino mixing and oscillation schemes with five neutrinos this problem of tension between data still exists.

 Many new  neutrino oscillation experiments designed  to check existing
indications in favor of short baseline neutrino oscillations  are  proposed  or  in preparation  at present (see recent review \cite{Lasserre:2014ita}). Proposed radioactive source experiments will be based on existing large detectors: Borexino \cite{Borexino:2013xxa}, KamLAND \cite{Gando:2013zoa},
Daya Bay \cite{Dwyer:2011xs,TheDayaBay:2013kda}. Important feature of these new experiments is a possibility to study $\frac{L}{E}$ dependence of $\nua{e}$ survival probability. Indications in favor of  the disappearance reactor $\bar\nu_{e}$'s will be checked in several future reactor neutrino experiments \cite{Lasserre:2014ita,Alekseev:2013dmu,Serebrov:2013yaa}. In these experiments spectral distortion as a function of the distance from reactor core will be studied. Anomaly, observed in the LSND experiment, will be investigated in future MiniBooNE+ experiment \cite{Dharmapalan:2013zcy}, in FermiLab experiment on the measurement of $\nu_{\mu}$ disappearance \cite{Anokhina:2014qda}, in ICARUS/NESSiE experiment \cite{Antonello:2013ypa,Rubbia:2014dva} with two LAr detectors. Direct test of the LSND anomaly is planned to be performed at the Spallation Neutron Source of the Oak Ridge Laboratory \cite{OscSNS:2013hua,Elnimr:2013wfa}. There exist a proposals to use for the search of sterile neutrinos muon storage ring, a source of $\nu_{e}$ and $\bar\nu_{\mu}$ (or  $\bar\nu_{e}$ and $\nu_{\mu}$)\cite{Adey:2014rfv}. There is no doubt that in a few years the problem of the existence of light sterile neutrinos will be fully clarified.

\begin{center}
\bf{On  the bariogenesis through leptogenesis }
\end{center}
Indirect indications in favor of existence of heavy Majorana leptons can be obtained from the cosmological data. From existing cosmological data it follows that our Universe predominantly consists of matter. For the barion-antibarion asymmetry we have
\begin{equation}\label{barantibar}
\eta_{B}=\frac{n_{B}-n_{\bar B}}{n_{\gamma}}\simeq \frac{n_{B}}{n_{\gamma}}=(6.11\pm0.19)\cdot 10^{-10}.
\end{equation}
Here $n_{B}$, $n_{\bar B}$ and $n_{\gamma}$ are barion, antibarion and photon number densities, respectively.

In the Standard Big Bang scenario initial numbers of barions and antibarions are equal. The observed barion-antibarion asymmetry have to be generated during the evolution of the Universe. A mechanism of the generation of the barion-antibarion asymmetry must satisfy the following Sakharov criteria \cite{Sakharov:1967dj}
\begin{enumerate}
  \item The barion number has to be violated at some stage of the evolution.
  \item $C$ and $CP$ must be violated.
  \item Departure from thermal equilibrium must take place.
\end{enumerate}
The interaction (\ref{HeavyMj}) with complex Yukawa couplings is a source of the $CP$ violation. Out of equilibrium $CP$ violating lepton-Higgs decays of heavy Majorana leptons, produced in the hot, expanding Universe,  could create lepton-antilepton asymmetry. This asymmetry, due to Standard Model nonperturbative sphaleron transitions in which $B$ and $L$ are violated, could be converted into barion-antibarion asymmetry (see reviews \cite{Buchmuller:2005eh,Strumia:2006qk,Davidson:2008bu,DiBari:2012fz}).

There are many models based on this general  scenario of bariogenesis through leptogenesis. {\em Existence of heavy Majorana leptons is their common feature.}

\section{Conclusion}

The Standard Model  successfully describes all observed physical phenomena in a wide range of energies up to a few hundreds GeV. After the discovery of the Higgs boson at LHC the Standard Model was established as  {\em a  theory of physical phenomena at the electroweak scale.} We suggest here that neutrinos play exceptional role in the Standard Model. Neutrinos apparently are crucial in the determination of symmetry properties of the Standard Model.

The Standard Model is based on
\begin{itemize}
  \item The local gauge symmetry.
  \item The unification of the weak and electromagnetic interactions.
  \item Brout-Englert-Higgs mechanism of the spontaneous breaking of the symmetry.
\end{itemize}
The Standard Model teaches us  that in the framework of these general principles nature choose the simplest possibilities. The simplest, most economical possibility for neutrinos is to be two-component  Weyl particles (Landau-Lee-Yang-Salam two-component neutrinos). The experiment showed that from two possibilities (left-handed or right-handed) nature choose the left-handed possibility.

In order to ensure symmetry, fields of quarks and leptons also must be two-component, left-handed and the symmetry group must be non-Abelian. This allow to include charged particles and ensure the universality of the minimal CC interaction of the fundamental fermions and the gauge fields. The simplest possibility is $SU_{L}(2)$ with doublets of the left-handed fields.

The unification of weak and electromagnetic interactions require enlargement of the symmetry group. The simplest possibility is $SU_{L}(2)\times U_{Y}(1)$ group. Because the electromagnetic current includes left-handed and right-handed fields of the {\em charged particles} charged right-handed fields  must be  SM fields (singlets of the $SU_{L}(2)$ group). Electric charges of neutrinos are equal to zero. The unification of the weak and electromagnetic interactions does not requires  right-handed neutrino fields. Minimal possibility is that {\em there are no right-handed neutrino fields in the SM.}  Nonconservation of $P$ and $C$ in the weak interaction apparently is connected with that. Because of there are no right-handed SM neutrino fields there is no Yukawa interaction which can generate neutrino mass term: neutrinos are the only particles which after spontaneous breaking of the electroweak symmetry  remain two-component left-handed.

With two-component left-handed neutrino fields $\nu_{lL}$ only a beyond the Standard Model, lepton number violating Majorana mass term can be built. {\em This is the most economical possibility.} It is generated by the unique, beyond the Standard Model dimension five Weinberg effective Lagrangian. Due to a suppression factor which is a ratio of the electroweak vacuum expectation value $v$ and the parameter $\Lambda$, which characterizes the scale of a new lepton number violating physics, such approach naturally explains the smallness of neutrino masses.

In the framework of the effective Lagrangian  values of neutrino masses, mixing angles and $CP$ phases can not be predicted. The same is true for leptons and quarks: the Higgs mechanism of the generation of masses and mixing of leptons and quarks do not predicts the values of masses, mixing angles and $CP$ phase. However, there are three general consequences of this mechanism of the neutrino mass generation.
\begin{enumerate}
  \item Neutrino with definite masses $\nu_{i}$ are Majorana particles.
  \item Number of neutrinos with definite masses is equal to the number of the flavor neutrinos (three).
 \end{enumerate}
The neutrino nature (Majorana or Dirac ?) can be inferred from the experiments
on the the search for neutrinoless double $\beta$-decay of $^{76} \mathrm{Ge}$, $^{136} \mathrm{Xe}$ and other nuclei. If this process will be observed it will be a proof that neutrinos with definite masses are Majorana particles, i.e. that neutrino masses have  a beyond the SM origin. Future experiments will probe inverted neutrino mass spectrum region ($m_{\beta\beta}\simeq \mathrm{a~few}10^{-2}$ eV).
In the case of normal mass hierarchy the probability of the neutrinoless double $\beta$-decay will be so small that new methods of the detection of the process must be developed (see \cite{Alonso:2014fwf}).

A possibility that the number of the neutrinos with definite masses is more than three will be tested in future
reactor, radioactive source and accelerator experiments on the search for sterile neutrinos.

The effective Lagrangian, responsible for the Majorana neutrino mass term, can be a result of the exchange of virtual heavy Majorana  leptons between lepton-Higgs pairs. The $CP$ violating, out of equilibrium decays of heavy Majorana leptons in the early Universe could be the origin of the barion-antibarion asymmetry of the Universe.

The value of the parameter $\Lambda$, which characterizes the scale of a new lepton-number violating physics, is an open problem. It is natural to assume that the Yukawa coupling constant is of the order of one. In this case $\Lambda\simeq 10^{15}$ GeV. However, much smaller values of $\Lambda$ can not be excluded. If $\Lambda$ is of the order of  TeV lepton-number violating decays of Majorana leptons can be observed at LHC (see, for example, \cite{Ibarra:2010xw,Strumia:2011zz,Molinaro:2013asa,Dev:2013oxa,Barry:2013xxa}).

The Standard Model teaches us that the simplest possibilities are more likely to be correct. Two-component left-handed Weyl neutrinos and absence of the right-handed neutrino fields in the Standard Model is the simplest, most elegant and most economical possibility. In this case generated by the effective, dimension five Lagrangian (or by the standard seesaw mechanism) {\em Majorana  mass term} (three Majorana neutrinos with definite masses, absence of sterile neutrinos) {\em is the simplest, most economical possibility.} Future experiments will show whether  this possibility is realized in nature.

This work is supported by the Alexander von Humboldt Stiftung, Bonn, Germany (contract Nr. 3.3-3-RUS/1002388),
by RFBR Grant N 13-02-01442 and by the Physics Department E15 of the Technical University Munich. I am thankful to W. Potzel for useful discussions and to the theory group of TRIUMF for the hospitality.


\begin{thebibliography}{99}

\bibitem{Fukuda:1998mi}
Super-Kamiokande Collaboration (Y.~Fukuda {\em et~al.}),
 {\em Phys. Rev.Lett.} {\bf 81}(1998) 1562, arXiv: hep-ex/9807003.


\bibitem{Ahmad:2002jz}
SNO Collaboration (Q.~R. Ahmad {\em et~al.}),
 {\em Phys. Rev. Lett.} {\bf 89} (2002) 011301, arXiv: nucl-ex/0204008.

\bibitem{Cleveland:1998nv}
Homestake Collaboration (B.~T. Cleveland {\em et~al.}),
 {\em Astrophys. J.} {\bf 496} (1998) 505.

\bibitem{Altmann:2005ix}
GNO Collaboration (M.~Altmann {\em et~al.}),
{\em Phys. Lett.} {\bf B616} (2005) 174, arXiv: hep-ex/0504037.

\bibitem{Abdurashitov:2002nt}
SAGE Collaboration (J.~N. Abdurashitov {\em et~al.}),
{\em J. Exp. Theor.Phys.} {\bf 95}(2002) 181, arXiv: astro-ph/0204245.






\bibitem{Araki:2004mb}
KamLAND Collaboration (T.~Araki {\em et~al.}),
{\em Phys. Rev. Lett.} {\bf94} (2005) 081801, arXiv:hep-ex/0406035.

\bibitem{Ahn:2006zza}
K2K Collaboration (M.~H. Ahn {\em et~al.}),
 {\em Phys. Rev.} {\bf D74} (2006) 072003, arXiv:hep-ex/0606032.



\bibitem{Adamson:2013whj}
MINOS Collaboration (P.~Adamson {\em et~al.}),
 {\em Phys.Rev.Lett.} {\bf 110}  (2013) 251801, arXiv:1304.6335 [hep-ex].

\bibitem{Abe:2014ugx}
T2K Collaboration (K.~Abe {\em et~al.}),
{\em Phys.Rev.Lett.} {\bf 112} (2014) 181801 , arXiv:1403.1532 [hep-ex].


\bibitem{An:2013zwz}
Daya Bay Collaboration (F.~An {\em et~al.}),
 {\em Phys.Rev.Lett.} {\bf 112} (2014) 061801, arXiv:1310.6732 [hep-ex].



\bibitem{Ahn:2012nd}
RENO Collaboration (S.-B. Kim {\em et~al.}),
{\em Phys. Rev. Lett.} {\bf 108}  (2012) 191802, arXiv:1204.0626 [hep-ex].





\bibitem{Abe:2013sxa}
Double Chooz Collaboration (Y.~Abe {\em et~al.}),
 {\em Phys.Lett.} {\bf B723},  (2013) 66, arXiv:1301.2948 [hep-ex].


\bibitem{Bellini:2014uqa}
BOREXINO Collaboration (G.~Bellini {\em et~al.}),
 {\em Nature} {\bf 512} (2014) 383.




\bibitem{Pontecorvo:1957cp}
B.~Pontecorvo,
{\em Sov. Phys. JETP} {\bf 6} (1957) 429, [Zh. Eksp. Teor. Fiz. 33, 549 (1957)].
\bibitem{Pontecorvo:1957qd}
B.~Pontecorvo,
{\em Sov. Phys. JETP} {\bf 7} (1958) 172, [Zh. Eksp. Teor. Fiz. 34, 247 (1958)].





\bibitem{Maki:1962mu}
Z.~Maki, M.~Nakagawa and S.~Sakata,
 {\em Prog. Theor. Phys.} {\bf 28} (1962) 870.














\bibitem{Bilenky:1987ty}
S.~M. Bilenky and S.~T. Petcov,
 {\em Rev. Mod. Phys.} {\bf 59} (1987) 671.


\bibitem{Bilenky:1998dt}
S. M. Bilenky, C. Giunti, W. Grimus,
Prog. Part. Nucl. Phys. \textbf{43} (1999) 1, hep-ph/9812360.


\bibitem{Bilenky:2011pk}
  S.~M.~Bilenky, F.~von Feilitzsch and W.~Potzel,
  J.\ Phys.\ G {\bf 38} (2011) 115002, arXiv:1102.2770 [hep-ph].

\bibitem{Bilenky:2012zp}
  S.~M.~Bilenky,
  arXiv:1208.2497 [hep-ph].


\bibitem{Bilenky:1980cx}
S.~M. Bilenky, J.~Hosek and S.~T. Petcov,
{\em Phys. Lett.} {\bf B94} (1980) 495.

\bibitem{Doi:1980yb}
M.~Doi, T.~Kotani, H.~Nishiura, K.~Okuda and E.~Takasugi,
 {\em Phys. Lett.} {\bf B102} (1981) 323.


\bibitem{Gonzalez-Garcia:2014bfa}
  M.~C.~Gonzalez-Garcia, M.~Maltoni and T.~Schwetz,
  JHEP {\bf 1411} (2014) 052, arXiv:1409.5439 [hep-ph].








\bibitem{Kraus:2004zw}
Ch. Kraus \textit{et al.},
Eur. Phys. J. \textbf{C40} (2005) 447, hep-ex/0412056.


\bibitem{Aseev:2011dq}
V.N. Aseev \textit{et al.} (Troitsk),
Phys. Rev. \textbf{D84} (2011) 112003, arXiv:1108.5034 [hep-ex].




\bibitem{Ade:2013zuv}
Planck Collaboration  (P.~A.~R.~Ade {\it et al.}),
 Astron.\ Astrophys.\  {\bf 571} (2014) A16, arXiv:1303.5076 [astro-ph].


















\bibitem{Minkowski:1977sc}
P.~Minkowski,
{\em Phys. Lett.} {\bf B67} (1977) 421.

\bibitem{Yanagida:1979as}
T.~Yanagida,
{\em Conf.Proc.} {\bf C7902131} (1979) 95.

\bibitem{GellMann:1980vs}
M.~Gell-Mann, P.~Ramond and R.~Slansky, {\em Conf.Proc.} {\bf C790927} (1979) 315, arXiv:1306.4669.

\bibitem{Glashow:1979nm}
S.~L. Glashow, {\em NATO Adv.Study Inst.Ser.B Phys.} {\bf 59} (1980) 687.

\bibitem{Mohapatra:1980ia}
R.~N. Mohapatra and G.~Senjanovic,
{\em Phys. Rev. Lett.} {\bf 44} (1980) 912.







\bibitem{Glashow:1961tr}
  S.~L.~Glashow,
  Nucl.\ Phys.\  {\bf 22} (1961) 579.




\bibitem{Weinberg:1967tq}
  S.~Weinberg,
  Phys.\ Rev.\ Lett.\  {\bf 19} (1967) 1264.

\bibitem{Salam:1968rm}
  A.~Salam,
  Conf.\ Proc.\ C {\bf 680519} (1968) 367.











\bibitem{Weyl} H. Weyl, 
Z. Physik  {\bf 56 } (1929) 330.




\bibitem{Pauli}W. Pauli, Handbuch der  Physik, Springer Verlag, Berlin   {\bf v.24 } (1933) 226-227.




\bibitem{Wu:1957my}
  C.~S.~Wu, E.~Ambler, R.~W.~Hayward, D.~D.~Hoppes and R.~P.~Hudson,
  Phys.\ Rev.\  {\bf 105} (1957) 1413.






\bibitem{Lederman} R. L. Garwin, L. M. Lederman and W. Weinrich,
Phys. Rev. {\bf 105} (1957) 1415.





\bibitem{Landau:1957tp}
  L.~D.~Landau,
  Nucl.\ Phys.\  {\bf 3} (1957) 127.

\bibitem{Lee:1957qr}
  T.~D.~Lee and C.~N.~Yang,
  Phys.\ Rev.\  {\bf 105} (1957) 1671.

\bibitem{Salam:1957st}
  A.~Salam,
  Nuovo Cim.\  {\bf 5} (1957) 299.









\bibitem{Goldhaber:1958nb}
  M.~Goldhaber, L.~Grodzins and A.~W.~Sunyar,
  Phys.\ Rev.\  {\bf 109} (1958) 1015.




\bibitem{Danby:1962nd}
  G.~Danby, J.~M.~Gaillard, K.~A.~Goulianos, L.~M.~Lederman, N.~B.~Mistry, M.~Schwartz and J.~Steinberger,
  Phys.\ Rev.\ Lett.\  {\bf 9} (1962) 36.

\bibitem{Kodama:2000mp}
DONUT Collaboration (K.~Kodama {\it et al.}),
  Phys.\ Lett.\ B {\bf 504} (2001) 218, hep-ex/0012035.


\bibitem{Englert:1964et}
  F.~Englert and R.~Brout,
  Phys.\ Rev.\ Lett.\  {\bf 13} (1964) 321.


\bibitem{Higgs:1964ia}
  P.~W.~Higgs,
  Phys.\ Lett.\  {\bf 12} (1964) 132.




\bibitem{Higgs:1964pj}
  P.~W.~Higgs,
  Phys.\ Rev.\ Lett.\  {\bf 13} (1964) 508.


\bibitem{Nambu:1960xd}
  Y.~Nambu,
  Phys.\ Rev.\ Lett.\  {\bf 4} (1960) 380.


\bibitem{Nambu:1961fr}
  Y.~Nambu and G.~Jona-Lasinio,
  Phys.\ Rev.\  {\bf 124} (1961) 246.


\bibitem{Goldstone:1961eq}
  J.~Goldstone,
  Nuovo Cim.\  {\bf 19} (1961) 154.



\bibitem{Aad:2012tfa} ATLAS Collaboration (G. Aad {\em et al.})
  Phys.\ Lett.\ B {\bf 716} (2012) 1, arXiv:1207.7214 [hep-ex].

\bibitem{Chatrchyan:2012ufa}
CMS Collaboration ( S.~Chatrchyan {\it et al.})  ,
  Phys.\ Lett.\ B {\bf 716} (2012) 30, arXiv:1207.7235 [hep-ex].









\bibitem{Beringer:1900zz}
Particle Data Group Collaboration ( J.~Beringer {\it et al.}) ,
  Phys.\ Rev.\ D {\bf 86} (2012) 010001.








\bibitem{Chatrchyan:2013zna}
 CMS Collaboration ( S.~Chatrchyan {\it et al.} ),
  Phys.\ Rev.\ D {\bf 89} (2014) 012003, arXiv:1310.3687 [hep-ex].

\bibitem{Chatrchyan:2014nva}
CMS Collaboration (S.~Chatrchyan {\it et al.}) ,
  JHEP {\bf 1405} (2014) 104, arXiv:1401.5041 [hep-ex].












\bibitem{Mohapatra:2006gs}
  R.~N.~Mohapatra and A.~Y.~Smirnov,
  Ann.\ Rev.\ Nucl.\ Part.\ Sci.\  {\bf 56} (2006) 569, hep-ph/0603118.



\bibitem{Weinberg:1979sa}
  S.~Weinberg,
  Phys.\ Rev.\ Lett.\  {\bf 43} (1979) 1566.


\bibitem{Wilczek:1979hc}
  F.~Wilczek and A.~Zee,
  Phys.\ Rev.\ Lett.\  {\bf 43} (1979) 1571.

\bibitem{Weinberg:1980bf}
  S.~Weinberg,
  Phys.\ Rev.\ D {\bf 22} (1980) 1694.


\bibitem{Gribov:1968kq}
  V.~N.~Gribov and B.~Pontecorvo,
  Phys.\ Lett.\ B {\bf 28} (1969) 493.



\bibitem{Doi:1985dx}
  M.~Doi, T.~Kotani and E.~Takasugi,
  Prog.\ Theor.\ Phys.\ Suppl.\  {\bf 83} (1985) 1.









\bibitem{Avignone:2007fu}
  F.~T.~Avignone III, S.~R.~Elliott and J.~Engel,
  Rev.\ Mod.\ Phys.\  {\bf 80} (2008) 481, arXiv:0708.1033 [nucl-ex].


\bibitem{Bilenky:2014uka}
  S.~M.~Bilenky and C.~Giunti,
  arXiv:1411.4791 [hep-ph].

\bibitem{Ryan-Okubo-NCS-2-234-1964}
C.~Ryan and S.~Okubo, {\em Nuovo Cimento Suppl.} {\bf 2} (1964) 234 .


\bibitem{Case:1957zza}
  K.~M.~Case,
  Phys.\ Rev.\  {\bf 107} (1957) 307.





\bibitem{Albert:2014awa}
J.B. Albert \textit{et al.} (EXO-200),
Nature \textbf{510} (2014) 229, arXiv:1402.6956 [nucl-ex].




\bibitem{TheKamLAND-Zen:2014lma}
The KamLAND-Zen Collaboration,
  arXiv:1409.0077 [physics.ins-det].






\bibitem{Agostini:2013mzu}
GERDA Collaboration (M. Agostini \textit{et al.}),
Phys.Rev.Lett. \textbf{111} (2013) 122503, arXiv:1307.4720 [nucl-ex].




\bibitem{KlapdorKleingrothaus:2004wj}
H.~Klapdor-Kleingrothaus, I.~Krivosheina, A.~Dietz and O.~Chkvorets, {\em Phys.
Lett.} {\bf B586} (2004)  198, hep-ph/0404088.



\bibitem{Klapdor-Kleingrothaus:2001yx}
H. V. Klapdor-Kleingrothaus \textit{et al.},
Eur. Phys. J. \textbf{A12} (2001) 147.


\bibitem{Aalseth:2002rf}
IGEX Collaboration (C. E. Aalseth \textit{et al.}),
Phys. Rev. \textbf{D65} (2002) 092007, hep-ex/0202026.


\bibitem{Aguilar:2001ty}
 LSND Collaboration ( A.~Aguilar-Arevalo {\it et al.}),
  Phys.\ Rev.\ D {\bf 64} (2001) 112007, hep-ex/0104049.

\bibitem{Aguilar-Arevalo:2013pmq}
MiniBooNE Collaboration (A.A. Aguilar-Arevalo \textit{et al.}),
Phys.Rev.Lett. \textbf{110} (2013) 161801, arXiv:1303.2588 [hep-ex].



\bibitem{Conrad:2013mka}
J. M. Conrad, W. C. Louis, M. H. Shaevitz,
Ann.Rev.Nucl.Part.Sci. \textbf{63} (2013) 45, arXiv:1306.6494 [hep-ex].



\bibitem{Mention:2011rk}
  G.~Mention, M.~Fechner, T.~Lasserre, T.~A.~Mueller, D.~Lhuillier, M.~Cribier and A.~Letourneau,
  Phys.\ Rev.\ D {\bf 83} (2011) 073006, arXiv:1101.2755 [hep-ex].




\bibitem{Mueller:2011nm}
Th. A. Mueller \textit{et al.},
Phys. Rev. \textbf{C83} (2011) 054615, arXiv:1101.2663 [hep-ex].

\bibitem{Huber:2011wv}
P. Huber,
Phys. Rev. \textbf{C84} (2011) 024617, arXiv:1106.0687 [hep-ph].

\bibitem{Kaether:2010ag}
  F.~Kaether, W.~Hampel, G.~Heusser, J.~Kiko and T.~Kirsten,
  Phys.\ Lett.\ B {\bf 685} (2010) 47, arXiv:1001.2731 [hep-ex].

\bibitem{Abdurashitov:2009tn}
SAGE Collaboration (  J.~N.~Abdurashitov {\it et al.}),
  Phys.\ Rev.\ C {\bf 80} (2009) 015807, arXiv:0901.2200 [nucl-ex].

\bibitem{Giunti:2013aea}
C. Giunti, M. Laveder, Y.F. Li, H.W. Long,
Phys.Rev. \textbf{D88} (2013) 073008, arXiv:1308.5288 [hep-ph].

\bibitem{Kopp:2013vaa}
J. Kopp, P. A. N. Machado, M. Maltoni, T. Schwetz,
JHEP \textbf{1305} (2013) 050, arXiv:1303.3011 [hep-ph].

\bibitem{Dydak:1983zq}
  F.~Dydak, G.~J.~Feldman, C.~Guyot, J.~P.~Merlo, H.~J.~Meyer, J.~Rothberg, J.~Steinberger and H.~Taureg {\it et al.},
  Phys.\ Lett.\ B {\bf 134} (1984) 281.


\bibitem{Adamson:2011ch}  MINOS Collaboration (P.~Adamson {\it et al.}),
  Phys.\ Rev.\ D {\bf 84} (2011) 071103, arXiv:1108.1509 [hep-ex].






\bibitem{Cheng:2012yy} MiniBooNE and SciBooNE Collaborations (G.~Cheng {\it et al.} ),
  Phys.\ Rev.\ D {\bf 86} (2012) 052009, arXiv:1208.0322.


\bibitem{Lasserre:2014ita}
T. Lasserre,
arXiv:1404.7352 [hep-ex],
13th International Conference on Topics in Astroparticle and Underground Physics, TAUP 2013.


\bibitem{Borexino:2013xxa}
Borexino Collaboration (G. Bellini \textit{et al.}),
JHEP \textbf{1308}, 038 (2013), arXiv:1304.7721 [physics].


\bibitem{Gando:2013zoa}
A. Gando \textit{et al.},
arXiv:1312.0896 [physics].


\bibitem{Dwyer:2011xs}
D.A. Dwyer, K.M. Heeger, B.R. Littlejohn, P. Vogel,
Phys.Rev. \textbf{D87}, 093002 (2013), arXiv:1109.6036 [hep-ex].

\bibitem{TheDayaBay:2013kda}
Daya Bay Collaboration,
arXiv:1309.7961 [hep-ex].


\bibitem{Alekseev:2013dmu}
V. Belov \textit{et al.},
Phys.Part.Nucl.Lett. \textbf{11}, 473 (2014), arXiv:1304.3696 [physics].



\bibitem{Serebrov:2013yaa}
A. P. Serebrov \textit{et al.},
arXiv:1310.5521 [physics].






\bibitem{Dharmapalan:2013zcy}
R. Dharmapalan \textit{et al.} (MiniBooNE+),
arXiv:1310.0076 [hep-ex].

\bibitem{Anokhina:2014qda}
A. Anokhina \textit{et al.},
arXiv:1404.2521 [hep-ph].





\bibitem{Antonello:2013ypa}
M. Antonello \textit{et al.},
arXiv:1312.7252 [physics].

\bibitem{Rubbia:2014dva}
C. Rubbia,
arXiv:1408.6431 [physics].








\bibitem{OscSNS:2013hua}
W. Louis \textit{et al.} (OscSNS),
arXiv:1305.4189 [hep-ex].




\bibitem{Elnimr:2013wfa}
M. Elnimr \textit{et al.} (OscSNS),
arXiv:1307.7097 [physics].




\bibitem{Adey:2014rfv}
 nuSTORM Collaboration ( D.~Adey {\it et al.})
  Phys.\ Rev.\ D {\bf 89} (2014) 071301, arXiv:1402.5250 [hep-ex].


\bibitem{Sakharov:1967dj}
  A.~D.~Sakharov,
  Pisma Zh.\ Eksp.\ Teor.\ Fiz.\  {\bf 5} (1967) 32
   [JETP Lett.\  {\bf 5} (1967) 24]
   [Sov.\ Phys.\ Usp.\  {\bf 34} (1991) 392]
   [Usp.\ Fiz.\ Nauk {\bf 161} (1991) 61].



\bibitem{Buchmuller:2005eh}
  W.~Buchmuller, R.~D.~Peccei and T.~Yanagida,
  Ann.\ Rev.\ Nucl.\ Part.\ Sci.\  {\bf 55} (2005) 311, hep-ph/0502169.


\bibitem{Strumia:2006qk}
  A.~Strumia,
hep-ph/0608347.

\bibitem{Davidson:2008bu}
  S.~Davidson, E.~Nardi and Y.~Nir,
  Phys.\ Rept.\  {\bf 466} (2008) 105, arXiv:0802.2962 [hep-ph].

\bibitem{DiBari:2012fz}
  P.~Di Bari,
  Contemp.\ Phys.\  {\bf 53} (2012) 4,  315, arXiv:1206.3168 [hep-ph].


\bibitem{Alonso:2014fwf}
  J.~R.~Alonso, N.~Barros, M.~Bergevin, A.~Bernstein, L.~Bignell, E.~Blucher, F.~Calaprice and J.~M.~Conrad {\it et al.},
  arXiv:1409.5864 [physics.ins-det].


\bibitem{Ibarra:2010xw}
  A.~Ibarra, E.~Molinaro and S.~T.~Petcov,
  JHEP {\bf 1009} (2010) 108, arXiv:1007.2378 [hep-ph].



\bibitem{Strumia:2011zz}
  A.~Strumia,
  PoS EPS {\bf -HEP2011} (2011) 098.

\bibitem{Molinaro:2013asa}
  E.~Molinaro,
  PoS EPS {\bf -HEP2013} (2013) 303, arXiv:1312.5076 [hep-ph].

\bibitem{Dev:2013oxa}
  C.~H.~Lee, P.~S.~Bhupal Dev and R.~N.~Mohapatra,
  Phys.\ Rev.\ D {\bf 88} (2013) 9,  093010, arXiv:1309.0774 [hep-ph].


\bibitem{Barry:2013xxa}
  J.~Barry and W.~Rodejohann,
  JHEP {\bf 1309} (2013) 153, arXiv:1303.6324 [hep-ph].

















\end{thebibliography}
\end{document}